\documentclass[twocolumn, amsmath,amssymb,superscriptaddress]{revtex4-1} 

\usepackage{dcolumn}
\usepackage{bm}
\usepackage{amsmath}
\usepackage{amsfonts}
\usepackage{graphics} 
\usepackage[dvips]{graphicx}
\usepackage{times}
\usepackage{setspace}
\usepackage{color}   
\usepackage{verbatim}
\usepackage[raggedright]{titlesec}
\usepackage[normalem]{ulem}
\usepackage{framed}

\usepackage{enumerate}
\usepackage[breaklinks=true]{hyperref} 
\usepackage{breakurl} 

\usepackage{balance}

\usepackage{dcolumn}
\newcolumntype{d}[1]{D{.}{.}{#1}}

\usepackage{lineno} 
\modulolinenumbers[2]                                                          

\hypersetup{colorlinks=true,citecolor=blue, linkcolor=black,urlcolor=blue} 
\usepackage[outercaption]{sidecap} 

\newcommand{\beginsupplement}{%
	\setcounter{page}{1}
        \setcounter{section}{0} 
        \renewcommand{\thesection}{S\arabic{section}}%
        \setcounter{table}{0}
        \renewcommand{\thetable}{S\arabic{table}}%
        \setcounter{figure}{0}
        \renewcommand{\thefigure}{S\arabic{figure}}%
        \setcounter{equation}{0}
        \renewcommand{\theequation}{S\arabic{equation}}%
     }
 
\begin{document}
\title{Shift  in house price estimates during COVID-19  \\ reveals  effect of crisis on collective speculation}
\author{Alexander Michael Petersen}
\affiliation{Department of Management of Complex Systems, Ernest and Julio Gallo Management Program, School of Engineering, University of California, Merced, California 95343, USA}

\begin{abstract}
\noindent We exploit a  city-level panel  comprised of individual house price estimates to estimate the impact of  COVID-19    on    both small and big real-estate markets in California USA. Descriptive analysis of spot house price estimates, including contemporaneous   price uncertainty and 30-day  price change for individual properties listed on the online   real-estate platform Zillow.com, together facilitate quantifying both the excess valuation and valuation confidence  attributable to this global socio-economic shock. Our quasi-experimental pre-/post-COVID-19 design spans several years around 2020 and leverages contemporaneous price estimates of rental properties -- i.e., real estate entering the habitation market,  just not for purchase (off-market) and hence free of speculation --  as an appropriate counterfactual to properties listed for sale,  which are subject to on-market  speculation. Combining unit-level matching  and  multivariate difference-in-difference regression approaches, we obtain consistent estimates regarding the sign and magnitude of  excess price growth observed after the pandemic onset. Specifically, our results indicate that properties listed for sale appreciated an additional  1\% per month above what would be expected in the absence of the pandemic. This corresponds to an excess annual price growth of  roughly 12.7 percentage points, which  accounts for more than half of the actual annual  price growth in 2021 observed across the studied  regions.
Simultaneously, uncertainty in price estimates decreased, signaling the irrational confidence characteristic of prior asset bubbles.  We explore  how these two trends are related to market size, local market supply and  borrowing costs, which altogether lend support for the counterintuitive roles of uncertainty and interruptions in  decision-making.  
\end{abstract}

\maketitle



 One of the most impactful financial life-course events that individuals may encounter is buying a house, 
and in the United States (US) this fundamental decision is increasingly facilitated by online real-estate platforms such as Zillow.com, Trulia.com and Redfin.com. These marketplace service platforms aggregate available property information  into  virtual marketplaces, thereby facilitating the rapid and remote comparison of individual candidate houses,  estimation of  mortgage repayment schedules,  and assessment of the overall real-estate market. Their user bases are broad, including  professional investors, traditional homeowners and sellers, and casual browsers alike \cite{ZillowPorn}.  
Consequently, the  inflow of high-frequency market information  that is aggregated by online   real-estate platforms  informs  potential buyer and seller speculation, defined as near-term expectations of price and price movements \cite{malpezzi2005role}, which is invariably conditioned by individuals varying and acutely sensitive perceptions of uncertainty.

\begin{figure*}
\centering{\includegraphics[width=0.99\textwidth]{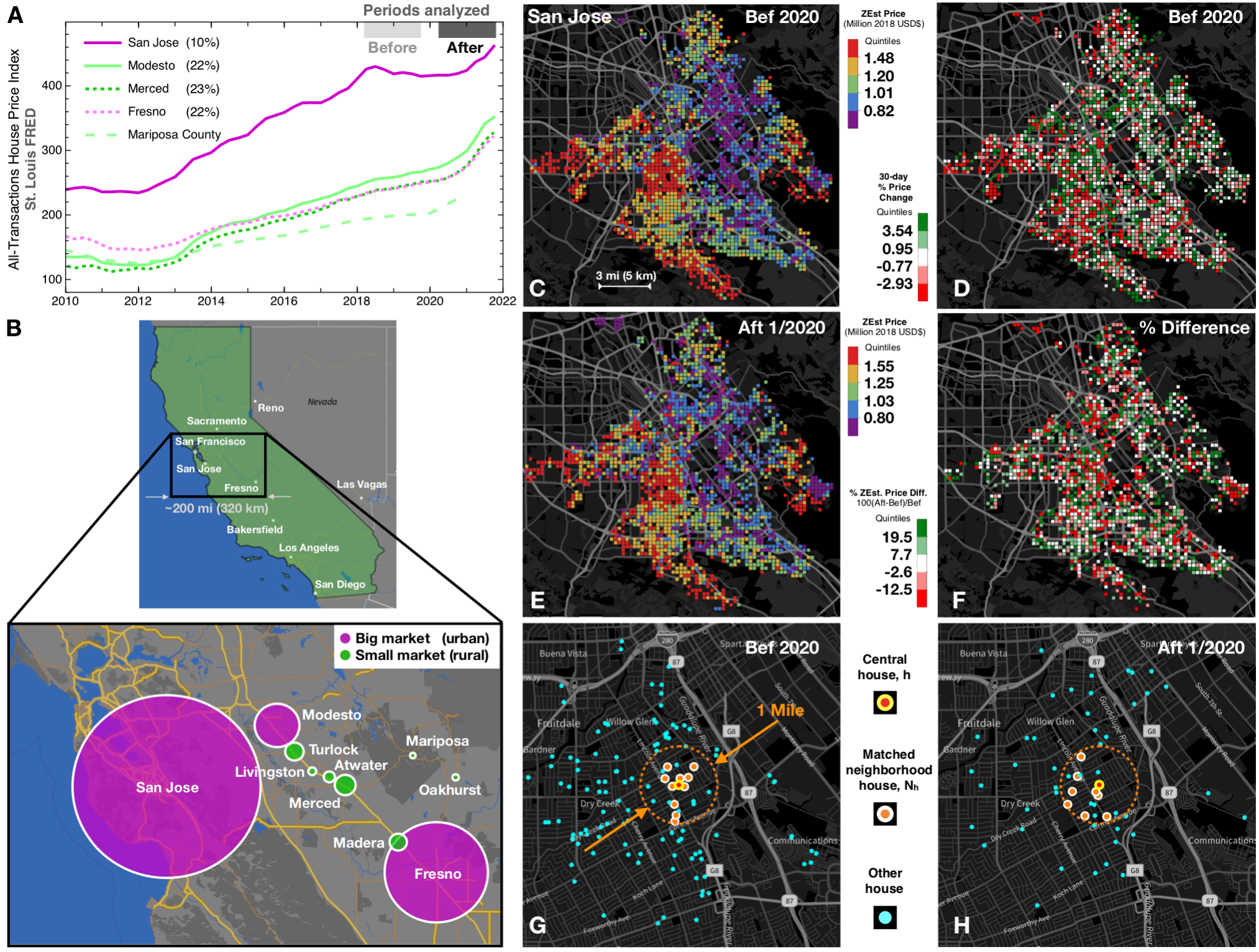}} 
 \caption{  \label{Fig1.fig} {\bf Schematic of data sampling and before- and after-1/2020 matching design.} 
 (A) All-Transactions House Price Index data by region,  obtained from the US Federal Reserve Bank of St. Louis (www.fred.stlouisfed.org). Annual percent increase from Oct. 2020-2021 are listed in the legend (2021 data not yet available for Mariposa; for more details see  {\bf Fig. S1}).  
(B) Longitudinal panel of Zillow Inc. house listings across 10 regions in northern California, USA constructed over  4-year time period  2018-2021   (see  {\bf Fig. S2} for sample size information). Shown are the locations and names of the 10 principal cities -- separated into big market (magenta) and small market (green) groups based upon 2021 population sizes, which are  proportional  to each  circle radius.  
(C) Spatial distribution  of mean house price estimate calculated for  properties listed for sale in San Jose before 2020; each grid is color-coded according to its corresponding distribution quintile.
(D) Mean 30-day price changes, color scale corresponds to distribution quintiles. 
(E) Mean price estimate  after 1/2020 using values deflated to 1/1/2018 US\$.
(F) Percent difference between grid values in panels B and D.
(G,H) Schematic of house matching design. For each  house listed  after 1/2020  (denoted by the index  $h$), we identified two  sets  of similar houses, denoted by  $\{N_{h}\}_{\text{Bef}}$ and $\{N_{h}\}_{\text{Aft}}$, based upon three criteria. Matched houses must be listed for sale in the same calendar month phase (e.g. if $h$ is from July then matches must be from May, June or July),  in the same price strata (i.e., matches must be within $\pm$ 1 price decile of $h$), and within a 1/2 mile radius of the central house. The set of matches $\{N_{h}\}_{\text{Bef}}$ are used for causal inference by way of a difference-in-difference identification strategy. The set  $\{N_{h}\}_{\text{Aft}}$ is only used to estimate the contemporaneous neighborhood housing supply, denoted by the activity $A_{h,m} = \vert \{N_{h}\}_{\text{Aft}}\vert$.  (G) Candidate matches before 2020 (10 matches indicated by orange dots); and (H) after 1/2020 (8 matches).  Candidate houses within the same period not meeting these criteria are indicated by blue dots. }
\end{figure*}

Against this backdrop, one of the many perplexing  outcomes of the  COVID-19 pandemic was the emergence of exuberant  markets in the US after the dust settled from the first shock wave.
This was particularly evident in the housing market, as illustrated in {\bf Fig. \ref{Fig1.fig}}(A), which shows the official  US government All-transactions House Price Index for several regions in  California (CA), where average home sale prices grew by up to 23\%  in 2021. Similar levels of   price appreciation  occurred in metropolitan areas across the US.

The initial reaction of US financial  and  housing markets to the COVID-19 outbreak   were sharply  negative, as this pervasive shock disrupted the health and security of individuals, thereby extending to  entire socio-economic systems   \cite{bonaccorsi2020economic,bunn2021covid,meyer2022pandemic,aleta2022quantifying}.
So why the rapid turnaround in these markets in the second half of 2020? 
Prior  empirical and theoretical work on real-estate markets  establishes various   factors underlying   market volatility, but distinct differences  in situational context make it challenging to infer wether or not the  prior insights readily extend to the events defining 2020-2021.
One particular factor unique to the pandemic period were stay-at-home orders that promptly and unexpectedly thrust an entire society into interpersonal interaction and information consumption modes that were entirely mediated by  the internet and electronic displays, the impacts of which are only now beginning to be understood \cite{liu2021covid}. This situational context is  relevant to our study given the prevalence and behavioral impact of online real-estate platforms in the US \cite{ZillowPopularityRank1,ZillowCoverage,ZillowPorn,SNLZillowPorn}, which establishes the conditions for multi-scale correlated phenomena that underly collective herding behavior \cite{sornette2017stock,mantegna1999introduction,roehner2002patterns,hong2008advisors,glaeser2008housing,shiller2015irrational}.    
Other relevant factors include the rapid deployment of work-from-home accommodations  that decreased the demand for metropolitan  amenities \cite{liu2021impact}, and also shifted   perspectives on work-life balance and associated household expenditures \cite{galanti2021work,gamber2021stuck}.

Understanding  the housing market's  response to macro-economic shock is critical to understanding the resilience of this fundamental global market. However, unlike stock markets, where an abundance of high-frequency data  provides a clear avenue for analyzing   market response to both anticipated and surprise news \cite{petersen2010quantitative,petersen2010market},  there are scant high-frequency   data sources for operationalizing such research on the real-estate market, even during  `normal' market periods. In this regard, our  data collection approach exhibits the utility of novel high-resolution and real-time altmetrics for research at  the intersection of  real estate and urban development \cite{steentoft2018canary,kaufmann2022scaling,pangallo2018home,bricongne2023web,fu2022human,seresinhe2016quantifying,botta2021modelling}.

In particular, we contribute to the literature on real-estate market dynamics and speculation  by tracking individual property valuations for nearly 2 years before and two years after the onset of the pandemic in January 2020, which we hereafter denote by ``1/2020''. 
A distinguishing feature of our study  is the construction of a high-resolution property-level dataset that captures two specific elements necessary for analyzing price speculation: (a) the 30-day change in estimated house price, which measures near-term price movements; and  (b)  the high-low range in   estimated house price, which  quantifies  uncertainty in price expectations.

As such, our multi-year analysis  leverages the sudden emergence of widespread  uncertainty as an instrument for analyzing the impact of collective speculation.  We leverage this systematic market shift by implementing a  difference-in-difference research design that compares price dynamics for properties listed for sale (on-market)  to  properties listed for rent (off-market) from the same neighborhood. These matched rental properties that were simultaneously available -- just not for sale, and thus transparent to  speculation deriving from short-term expectations of  returns via resale --  provide a counterfactual   baseline  for estimating  our main result: to what degree was excess  real-estate price growth  attributable to  COVID-19 pandemic uncertainty? 

In what follows we  address this question by way of  the  following  three research questions.
First, what are the  characteristics of high-frequency real estate price dynamics at the 1-month resolution, and to what degree did they  change  after the  COVID-19 pandemic? Second, to what degree did the pandemic shock to market uncertainty affect collective speculation -- namely, in  house price estimates and certainty in those estimates? And third, how did shifts in speculation  relate to fundamental market factors, such as market size, supply, and benchmark borrowing rates? While our results are based upon regions in California, our results  are likely generalizable to  similar US regions featuring speculative growth and subsequent price relaxation (see \href{https://twitter.com/NewsLambert/status/1602336916027981824}{``Shift in home prices since their respective 2022 peak''} (L. Lambert, Fortune.com); corresponding full article published by \href{https://fortune.com/2022/12/03/housing-market-moodys-updated-home-price-correction-forecast-housing-crash/?queryly=related_article}{Fortune.com}) given the ubiquity of the underlying market factors (low interest rates, high uncertainty, supply constraints)   during our sample period. 

\vspace{-0.3in}
\section*{Literature review}
\vspace{-0.2in}
This work contributes to two distinct research streams. First, the empirical analysis of  real-estate price dynamics, price elasticity \cite{saiz2010geographic,landvoigt2015housing,defusco2017interest,baum2019microgeography} and the overall real-estate market's response to exogenous shocks \cite{kaplan2020housing,d2022covid,ramani2021donut,Zillow2_Petersen_2023}. 
And second,  the understanding of  decision-making under extreme uncertainty following sudden interruptions to normal daily life \cite{tormala2016role,kupor2015persuasion} within the context of the COVID-19 pandemic  \cite{bonaccorsi2020economic,bunn2021covid,meyer2022pandemic,aleta2022quantifying,conley2021past}.

A  common methodology in the real-estate literature are hedonic regression models, applied to  identify attributes associated with a given property and neighborhood  that are positively and negatively correlated with property valuations.  Hedonic factors include property-level features such as building type, materials and floor area,  combined with important local amenities \cite{kaufmann2022scaling} such as  access to public transportation \cite{andersson2010does,ibeas2012modelling}, and security of clean tap water  \cite{cho2011negative,theising2019lead,mamun2023valuing}.  Other studies identify  externalities that are pervasive, such as climate change impacts on tree shade coverage \cite{siriwardena2016implicit}.

We do not employ hedonic factor analysis in this work, because our data source lacks consistent property-level features. Moreover,  we do not  model the  estimated property valuation   nor its final sale price. 
 Instead, we take estimated  property  valuations as a given, and then analyze how  valuation changes are correlated with micro-economic factors such as market size, local housing supply, and benchmark borrowing rates.  
 
 According to established  economic theory, lower mortgage rates  contribute to increased housing demand \cite{landvoigt2015housing,HousingBoomMortgageRates}. 
Yet few housing market analyses are performed over periods featuring systematic urban-to-rural  migration, as observed in the US during the pandemic \cite{coven2022urban}, because  most studies focus principally on select large metropolitan markets. Hence, there is  scant research  comparing urban and rural markets within the same region and period. 
As such, a distinction of the present work  is the construction of a balanced panel of multiple neighboring regions, for both large and small market size,  over a significant time horizon. 
We selected the 10 locations analyzed hereafter based upon  the accessibility of consistent property-level data from Zillow.com, familiarity with the region, and most importantly,   regional context. Namely, California has been affected for decades by an affordable housing crisis  that is concentrated in regions with high wealth inequality, the Bay Area mega-region being a case example \cite{loftus2011tent,raetz2020hard}.

  Another issue that has limited research on real-estate market dynamics is the scant availability of high-resolution data at the property level. Instead, market research commonly employs  annualized property sales data that are  aggregated at the regional level, which fails to capture  market dynamics  of individual properties. 
This choice follows from the technical challenges associated with assembling a balanced panel comprised of data with high spatiotemporal  resolution  by sourcing data  from online real-estate platforms,  with few examples  \cite{fu2022human,bricongne2023web}. Instead, much of the related micro-economic  literature  uses  house sales transaction data aggregated as mean values over sizable regions such as US ZIP codes or census tracts 
\cite{Himmelberg_2005,landvoigt2015housing,baum2019microgeography,liu2021impact,d2022covid,gamber2021stuck}. 
One example is the recent work by Mondragon \& Wieland \cite{mondragon2022housing} who use  house  transaction data aggregated across US counties over the period 12/2018--11/2021, reporting that a 1\% increase in  a region's share of remote-work explains 0.93\% increase in average house prices across the US, which accounts for roughly half of the price growth over that period analyzed. 
  
As the unit of analysis in our study is an individual property, as opposed to the median or average property within a specific zip code or other regional unit, we also contribute to empirical research on micro-level asset price dynamics \cite{steentoft2018canary}. 
Various asset classes, such as  stock prices, firm  sizes and human productivity, are amenable to analysis over variable time windows ranging from intraday, to monthly, to intra-annual and decadal scales  \cite{mantegna1999introduction,plerou1999similarities,petersen2012persistence,buldyrev2020rise}. 
The most relevant study of real-estate market dynamics is by \cite{landvoigt2015housing}, who analyze capital gains on sold properties over a 5-year horizon for the specific region of San Diego, CA.
We are unaware of  research analyzing the dynamics of individual real-estate valuations at the 1-month frequency, which is a unique feature of our data source. 

A final consideration regarding the extant COVID-19 research is the predominant focus on the short-term market decline in real estate markets   immediately following the onset of the pandemic \cite{liu2021impact,balemi2021covid,ramani2021donut,d2022covid}. This focus neglects the overwhelming market reversal that followed the initial negative market reaction. Such a narrow window also  disregards the pre-existing trends in market appreciation that preceded the pandemic in California, USA and elsewhere. 

\vspace{-0.2in}
\section*{Methods}
\vspace{-0.2in}
 \noindent{\bf Data source.} 
\noindent We  constructed a balanced 10-region panel with four notable features. 
 First, we  collected   property-level data at  high spatiotemporal resolution from a prominent online  real-estate platform (Zillow.com).
  As such, the fundamental unit of observation in this balanced panel are individual house listings, which distinguishes our study from much of the prior  literature. 
In total, the dataset is comprised of  57,414 individual properties listings from 10 regions  spanning a nearly 4-year time period (2018-2021) \cite{DataDryad}. 
 Notably, we do not include off-market  properties (those that are not listed either for sale or rent on Zillow.com),  even though Zillow Inc. produces and updates property valuation estimates for all on and off-market properties within its massive and near comprehensive real-estate data for the US market. 

 {\bf Figure \ref{Fig1.fig}}(B) shows the location of the 10 regions, which are  official administrative units in  CA. Individual house-level data were collected from the official Zillow Inc. application programming interface (API). For each month (from March 2018 to September 2021) and each region,  we used the  open-access Zillow Inc. GetSearchResults  API to collect comprehensive data on all on-market properties belonging to either of  two property categories: ``For Sale'' and ``Rent''. 
 For  further elaboration on the  available house-level data see  the official Zillow API  page  \cite{ZillowAPI}. 
 
As such, because our panel  includes high variation in  region sizes and population density,  these data can be used to compare  market dynamics according to housing market size. Three regions are associated with big (urban) markets (San Jose, Modesto, Fresno), and the remaining seven are associated with small (rural) markets, as proxied by the  principal city population for each region. Because these regions all belong to the Bay Area mega-region, connected together by a major public highway, we are able to   estimate the differential impact of the pandemic on urban versus rural  settings within the same macro-region backdrop. This approach distinguishes our study from other studies that  focus on just the largest metropolitan real-estate markets.

Second,  as  the top real-estate website in the U.S. in 2021 with roughly  36 million visits per month  \cite{ZillowPopularityRank1}, Zillow Inc. is a leading real-estate platform in an increasingly ubiquitous   IT service sector \cite{maglio2019handbook}.  By  maintaining a nearly real-time catalogue of  available listings and estimated valuations,  Zillow  facilitates comprehensive market assessment in addition to mediating buyer-seller interactions. Consequently, data obtained from the Zillow API are algorithmically consistent, which is critical for  analyzing simultaneous snapshots of  entire  regional housing markets. Alternative methods  collecting ask and sales prices from regional multiple listing services (MLS) involve data collected from different brokers, realtors and sellers, and  do not satisfy this consistency criterion.

Third, our dataset includes  quantitative measures of   speculation and uncertainty within the real-estate asset class, for which little is known.
Specifically,  Zillow collects, integrates and calculates  real-time house price estimates, including a 30-day  price estimate change, along with high and low  price estimates for each property. These property valuations derive from a proprietary in-house algorithm that estimates individual house prices based upon a massive and near comprehensive historical database  extending back to the mid 2000s, including ask prices elected by the sellers and subsequent  sale prices. 
These primary source data are readily available to the public and have fostered data science education and research by way of  open  competitions \cite{ZillowAPI,ZillowKaggle}.

Zillow   house price estimates are not only calculated  at the point of market entry (typically when the seller declares a public ask price), but are also interpolated  between prior listing and future price updates  in real time. As such, even though Zillow price estimates are   algorithmically determined, they integrate contemporaneous macroeconomic,  regional, neighborhood and house-specific factors rendering the estimates consistent and robust. 
Moreover, price estimates are rapidly calibrated to property sale events -- not only of the individual property itself, but also its neighbors, which contributes to a collective mode of price formation and speculation \cite{fu2022human}.
This is in contrast  to non-centralized data sources such as Multiple Listing Service (MLS) databases, which aggregate  listing information that may depend upon  realtors' and owners' idiosyncratic understanding of price formation and speculation.  

A fourth   feature of  the data source  is the consistent property value estimation for properties listed for sale and for rent. 
In the present study, rental properties   entering the habitation market played an important  role in accommodating the desire to escape high population density and/or to take advantage of remote work opportunity -- two factors associated with the pandemic housing market. 
Hence, in what follows we juxtapose the price dynamics for these  two distinct classes of available real estate  to estimate the impact of pandemic speculation on the housing market.
The  key distinction  being that  buyer-seller interactions   implicitly incorporate speculation on future price movements. By contrast, rental property owners instead opt for a revenue strategy based upon cash flow derived from future rents, which is less dependent on property and real-estate market speculation. 
To be clear,    data obtained for  rental listings  are not monthly rent estimates, but  are estimated  valuations of the rental {\it property}, i.e. deriving from same algorithm  as those properties that are listed for sale, rendering these distinct property classes directly comparable. \\

 \noindent{\bf Data Collection.}
\noindent 
We obtained data for  10 proximal CA cities and their surrounding regions  belonging to the Bay Area mega-region  shown in {\bf Fig. \ref{Fig1.fig}}(B).  
The largest principal city by population  is San Jose   ($\sim$1 million inhabitants in 2021); and by area is Fresno  (116 square miles); the smallest city by population is Mariposa ($\sim$1500 inhabitants) and by area is Livingston (3.7 square miles). For spatiotemporal context, the distance separating San Jose and Fresno is roughly 150 driving miles (240 km) corresponding to 2.5 driving hours. Despite a wide variation in size, location and socio-economic backdrop, these 10 regions all feature  shortages in affordable housing, a longstanding problem plaguing California and various other metropolitan areas in the United States \cite{loftus2011tent,raetz2020hard}.  Seven of the principal cities are located along a major industrial and commuter transportation highway (CA 99), and are within the 3-hour super-commuter travel-time from the greater Bay Area, thereby qualifying as bedroom communities. 
Conversely, two regions (Mariposa and Oakhurst) are oriented around recreational tourism  in and around Yosemite National Park. 
All together, these municipalities span a  wide range of house prices, market size and turnover to support within and across-city analysis at high geo-temporal resolution. 

 In the remainder of the analysis, for  data sampled  between March 2018 and May 2019, we denote this sample as ``before 2020''; and for data sampled  between May 2020 and September 2021, we denote this sample as  ``after 1/2020''. 
See {\bf Fig. S2}(A,B) for  monthly sample sizes for data grouped by   property type (``For Sale'' and ``Rent''); and {\bf Fig. S2}(C,D) for sample sizes grouped by  6-month non-overlapping periods that facilitate a visual comparison of average-property trends before and after 1/2020.

Each month we first obtained a set of unique listing identifiers (ZPID) by manually scanning across the entire Zillow.com directory for a given region and property type. 
This sampling frequency is sufficient to collect data for the majority of listings made within a monthly time window, as the average property during this period was on the market for 44 days \cite{d2022covid}.
To ensure sample time consistency and to also be in in accordance with  daily API call limits \cite{ZillowAPI}, we  limited  sampling to just these 10 regions. 
Consequently, API requests spanned just a couple  days  each month, and are thus contemporaneously consistent. Notably, a listing is featured in either of the property type  catalogues at the owner's or realtor's discretion, and so we do not capture hidden or private listings, which is a   limitation to our approach. However, given the prominence of Zillow.com in the US \cite{ZillowPopularityRank1}, we believe  this sampling bias is very limited in scope.\\
 
\begin{SCfigure*}
\includegraphics[width=0.575\textwidth]{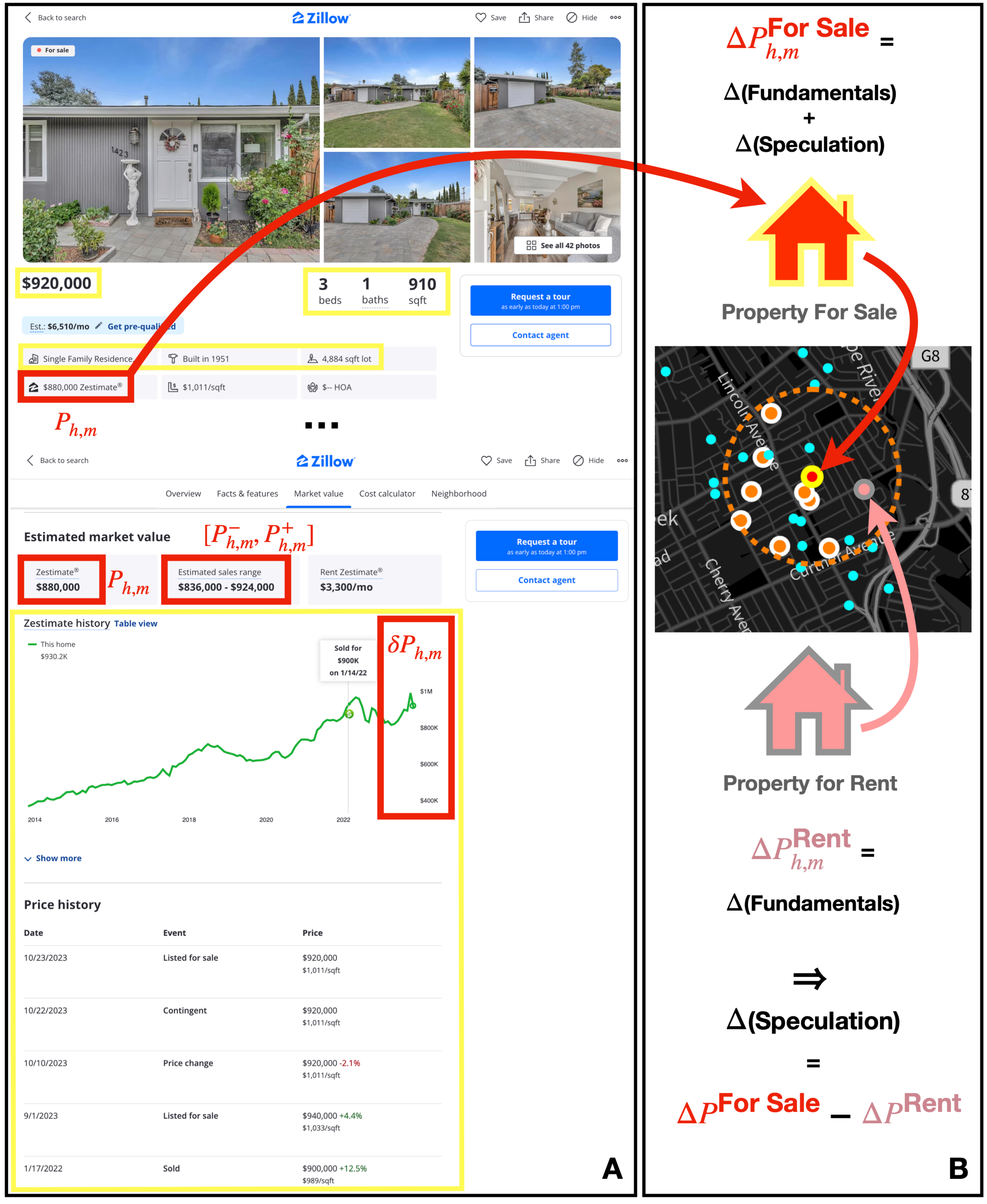}
 \caption{    \label{Fig2.fig} {\bf Schematic of quasi-experimental design for estimating the magnitude of price shifts attributable to COVID-19 market speculation.} 
  (A) Shown is a Zillow webpage for an actual on-market property listed for sale. Red highlights indicate the primary source data obtained from the open-access Zillow Inc. GetSearchResults  API; yellow highlights indicate additional  standardized data that feed into the proprietary Zillow Inc. algorithm that yields real-time  estimates for $P_{h}$, $\delta P_{h}$, $P^{+}_{h}$ and $P^{-}_{h}$. In addition to contemporaneous valuation estimates, users  are also confronted with longitudinal $P_{h}(t)$ histories extending up to a decade, which includes actual sales events  indicated in the ``Price History'' section of each listing page. (B) Our quasi-experimental design  leverages the algorithmically consistent data ($P_{h}$, $\delta P_{h}$, $P^{+}_{h}$ and $P^{-}_{h}$) available  for on-market properties listed for sale (which are sensitive to market speculation)  as well off-market properties listed  for rent.  Rental properties represent appropriate counterfactuals in that while they are available for habitation, they are off-market, meaning that they are neutral to short-term market speculation (since the time horizon for entering the market is well beyond the horizon for contemporaneous speculation). 
 Consequently, whereas price changes for on-market properties depend on shifts in the valuation of fundamentals  in addition to market speculation, price changes for rental properties primarily reflect  shifts in the valuation of fundamentals (e.g., the incremental  value of an additional bedroom). 
  Hence, this study applies a difference-in-difference (DiD) design to net out  shifts in the valuation of fundamentals in order to isolate    shifts  attributable to speculation -- see Eq. (\ref{EQDiD}). Moreover, by  comparing shifts after versus before 1/2020, we estimate the  effect of  market speculation deriving from COVID-19 uncertainty on the real-estate market.}
\end{SCfigure*}

\noindent{\bf Property-level metrics.}
The primary data used in this study come  from two open data sources: the US Federal Reserve Bank of St. Louis and  Zillow Inc.
From the US Federal Reserve we  collected  monthly data  compiled by Freddie Mac$^{\mbox{\scriptsize\textregistered}}$ for the \href{https://fred.stlouisfed.org/series/MORTGAGE30US}{average US 30-year fixed rate mortgage}, denoted by $M_{m}$, which provides a macro-economic indicator of borrowing costs. 
From Zillow Inc. we  exploit their internal system of unique property identifiers (ZPID) that  facilitate property  disambiguation  to assemble a city-level panel of  property-level data.  Specifically, for each unique property $h$ in  sampling month $m$,  we obtained the following data from the Zillow GetSearchResults API: 
\begin{enumerate}
\item the official address (including zip code and city name); 
\item the longitude and latitude (centroid of the property); 
\item the Zillow price estimate, termed the Zestimate$^{\mbox{\scriptsize\textregistered}}$, which we denote by $P_{h,m}$; 
\item the high and low range for $P_{h,m}$, denoted by $P^{+}_{h,m}$ and $P^{-}_{h,m}$, respectively; 
\item  the 30-day change in the $P_{h,m}$, denoted by $\delta P_{h,m}$. 
\end{enumerate}
{\bf Figure \ref{Fig2.fig}}(A) shows a sample Zillow webpage for a property in San Jose CA, illustrating the   prominence of  contemporaneous $P_{h}$, $\delta P_{h}$, $P^{+}_{h}$ and $P^{-}_{h}$ data as well as 10 years of historical data that confronts both casual and purposeful platform users.

The price estimates  ($P_{h,m}$ and $\delta P_{h,m}$)  are  calculated by Zillow Inc. based upon their proprietary in-house algorithm that  incorporates a  battery of   hedonic factors. For example,  inputs used to estimate $P_{h,m}$ include macro-economic market data (such as mortgage rates, regional and neighborhood data such as schools and similar houses),  house-specific data provided by the seller and from external sources (habitation area, number of floors, construction materials and date, pool and yard dimensions, garage capacity, school district, neighborhood amenities, and other  web-metrics such as house-views), and other properties in the neighborhood of $h$ that are either contemporaneously for sale or were listed in the past (i.e., within the near-comprehensive Zillow Inc. property database). 

Note that $P_{h,m}$ is  not  the asking price set by the listing agent, but rather an estimate of the property's market value. 
It is common for Zillow.com property profiles to feature up to 10 years of historical price estimates as a time series, also annotated by point events corresponding to prior ask and sales prices, which together inform buyer and seller speculation.  
Manual inspection of   10-year  Zestimate$^{\mbox{\scriptsize\textregistered}}$ time series indicates that new listings and updated  ask prices are rapidly  incorporated into the Zestimate$^{\mbox{\scriptsize\textregistered}}$ algorithm \cite{fu2022human}. This rapid information collection is a critical feature that facilitates collective co-production of market speculation deriving from individual seller and online platform service user activity.
In this regard, $P_{h,m}$ represents a dynamically updated estimate of the  fair market value based upon real-time, localized and comprehensive market information. 

 Notably, the Zestimate$^{\mbox{\scriptsize\textregistered}}$ error rate, measured as the percent difference between  $P_{h,m}$ and the property's actual sale price,  has decreased over time as their proprietary algorithm becomes more  accurate.
According to Zillow Inc., the median error rate (such that 50\% of property valuation errors are less than this value)  for on-market homes was 3.2\% during our sampling period, and has since decreased to 2.4\% \cite{ZillowZestimate}. 

These unique features of Zillow property data -- namely, the comprehensiveness, consistency,  dynamics and accuracy --  facilitate analyzing the evolution of the housing market in  specific regions at high geographic and temporal resolution. 
Without this rich data source, the next best alternative would be to pool records of  seller ask prices. However, such data would not be consistent and would not include dynamics, as the ask price occurs at a fixed date and does not tend to  change over a 30-day time window. Instead, the Zestimate$^{\mbox{\scriptsize\textregistered}}$ is updated in real time.  
Also, seller ask prices  do not include a price range, and so they do not permit analysis of valuation uncertainty.

We constructed our panel of  Zillow property estimates  by sampling Zillow.com  monthly for over 4 years. 
As such,  price values were obtained in nominal US\$ at the sampling month $m$. Hence, in what follows, we deflated all price values  to 1/1/2018 US\$.
We   control for the data sampling (calendar) month in our statistical analysis  to account for well-known intra-annual  housing market activity cycles  \cite{FirstTueCAMarket}.

Based upon the primary data from Zillow.com, we  also  computed three additional metrics. First, we calculated the price change as a percent of the initial price, 
\begin{eqnarray}  
\Delta P_{h,m} =100 \times \frac{\delta P_{h,m}}{P_{h,m}-\delta P_{h,m}} \ .
\end{eqnarray}   
See {\bf Fig. S2}(E,F) for the mean and standard deviation of  $\Delta P_{h,m}$,  grouped by period and property type.  Second, we calculated the  spot price  uncertainty, 
\begin{eqnarray}  
U_{h,m}=100 \times\frac{P^{+}_{h,m}-P^{-}_{h,m}}{P_{h,m}} \ .
\end{eqnarray}  
See {\bf Fig. S2}(G,H) for the mean and std. dev. of  $U_{h,m}$, grouped by period and property type.
And third, we  estimated the neighborhood housing market activity $A_{h,m}$ of a particular listing $h$ by counting the total number of properties  within a 0.5 mile (0.8 km) radius, and within the contemporanous three-month period $\{m_h, m_h-1,m_h-2\}$   including the listing month  $m_h$. 

 \noindent{\bf Pruned data sample optimized for unit-level matching.} Our quantitative analysis focuses on typical property listings for which there is sufficient neighborhood activity to support unit-level matching. For this reason we exclude observations from our raw data sample according to four criteria. 

First, we excluded property listings featuring extreme price change or price uncertainty values. Specifically, we only include properties with $\Delta P_{h,m} \leq 40 \%$ and  $U_{h,m} \leq 40 \%$, which together reduced the original dataset from 133,668 observations to 110,530 listings (a 17\% reduction). 
 
Second, to ensure properties have sufficient real-estate activity in the neighboring vicinity that offers alternative buyer options, we excluded  properties with fewer than 4  listings within the local neighborhood, defined as a 0.5 mile radius around $h$ -- which, for example, corresponds to  10 New York City blocks. This choice ensures that  comparable  properties used in our unit-level matching approach can be reached by walking, and so in principle have the same  neighborhood amenities as the central property.  

Third, we only consider alternative property listings from the same calendar phase, defined as  a three-month window prior to and including the central property's listing month. 
That is, if a property was listed in April 2021, we only consider candidate matches in the before 1/2020 period that were listed in February, March and April.
And fourth, we exclude listings outside of the active CA real-estate period, which is March thru October \cite{FirstTueCAMarket}. 

Together, the second, third and fourth stages of pruning   further reduced the sample size from 110,530 to 57,414 listings, corresponding to a 48\% reduction, largely attributable the second criteria regarding neighborhood activity. Together, these latter three criteria eliminated  many listings corresponding to empty lots and other under-developed properties located beyond the principal city limits associated with each region.
  
\begin{SCfigure*}
\includegraphics[width=0.69\textwidth]{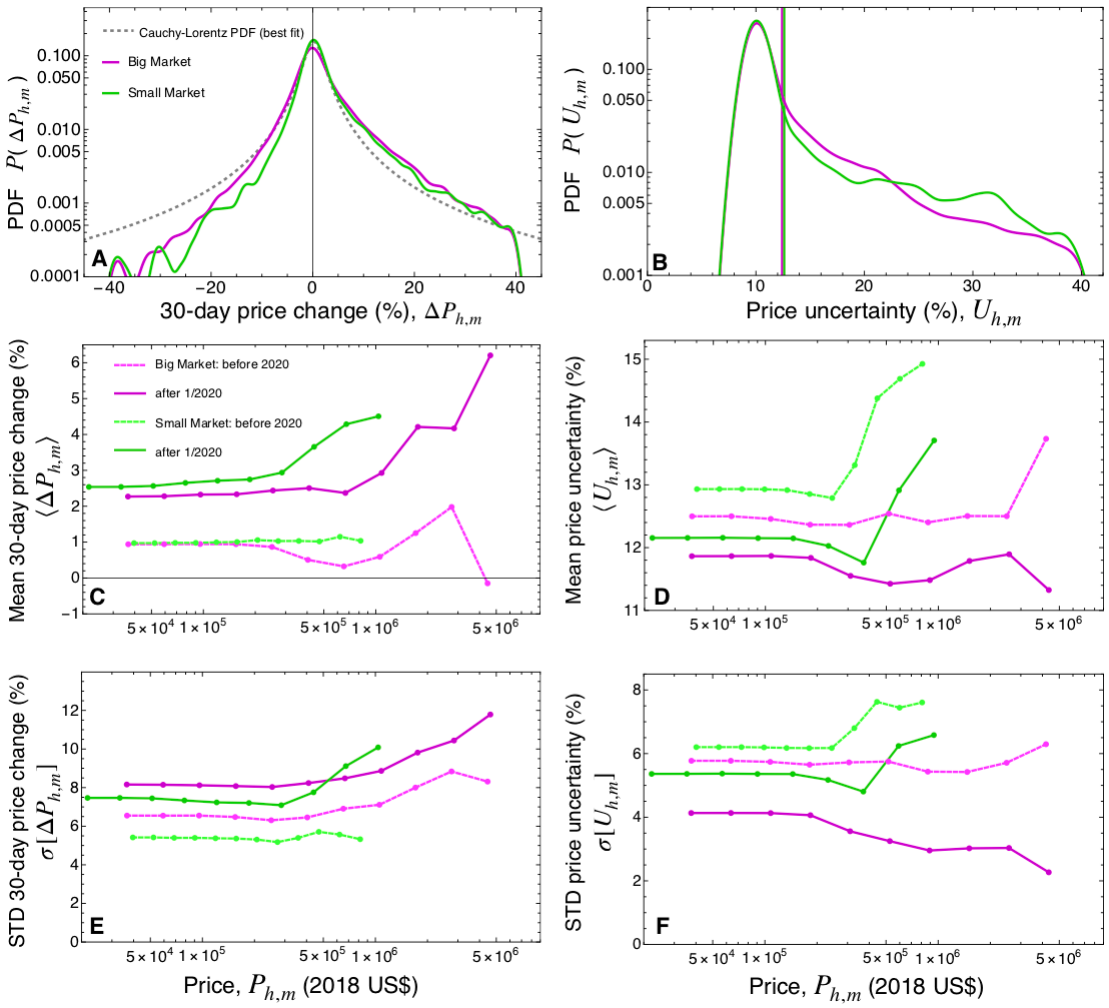}
 \caption{  \label{Fig3.fig} {\bf Systematic increase in property valuation and confidence in the after-1/2020 housing market.} 
Kernel density estimate of the probability density function (PDF) calculated for (A) 30-day price change, $\Delta P_{h,m}$, including  the best-fit Cauchy PDF calculated using both the big and small market data combined; and (B) PDF calculated for price uncertainty, $U_{h,m}$. Data shown are calculated using properties listed ``For Sale''; see {\bf Fig. S4} for PDF conditioned on market size, period and property type.
(C-F) Mean ($\langle \cdot \rangle$) and standard deviation (STD, $\sigma [\cdot]$)  calculated for $\Delta P_{h,m}$  and  $U_{h,m}$ conditional on spot price $P_{h,m}$. Together, these two variables show how the after-1/2020 CA housing market features  excess valuation growth  and  increasing valuation confidence (i.e., decreased uncertainty), patterns that are common to both the big and small markets, and appear to be even stronger for the small market. 
 These effects manifest as  systematic shifts in the first and second moments  --  i.e., the characteristic  location (C,D) and characteristic  fluctuation scale (E,F)  -- of the underlying data distributions, and are robust across the entire range of house listing price estimates.}
\end{SCfigure*}

\vspace{-0.3in}
\section*{Results}
\vspace{-0.2in}
\noindent{\bf Descriptive statistics grouped by region and period.}
The distribution of house prices $P_{h,m}$ is approximately log-normal -- see {\bf Fig. S3}. 
This feature is  consistent with the Gibrat proportional  growth model \cite{mantegna1999introduction,buldyrev2020rise}.

{\bf Figure \ref{Fig3.fig}}(A,B) show the distributions of $\Delta P_{h}$ and $U_{h}$, respectively.
30-day price-change fluctuations ($\Delta P_{h}$) feature high levels of variance around the roughly 1-2\% average price growth levels observed during the sample period. 
The frequency distribution $P(\Delta P_{h})$  is asymmetric and leptokurtic, being wider in the bulk than the Laplace (double-exponential) tent-shaped  growth  distributions observed in other relevant empirical studies of economic growth  \cite{stanley1996scaling,plerou1999similarities,buldyrev2020rise,petersen2012persistence}. Both the positive and negative tails of $P(\Delta P_{h})$  are heavy, extending well beyond the  values of $\pm 40\%$ used to truncate our data sample (a sampling choice used so  that parameter estimates in our regression model are not biased by extreme outliers). 

 We estimated a best model for the $P(\Delta P_{h})$ distribution   using the maximum likelihood method. 
The resulting best fit  probability density function (PDF)  is the Cauchy-Lorentz distribution,  
\begin{eqnarray}  
P(\Delta P) = \frac{1}{\pi \gamma \big(1+(\frac{\Delta P-x_{0}}{\gamma})^{2} \big) }\ ,
\end{eqnarray} 
which has asymptotic power-law tail behavior $P(\Delta P) \sim \Delta P^{-2}$ for $\vert \Delta P - x_{0}\vert \gg \gamma $. 
 The two Cauchy-Lorentz PDF parameters  estimated   using both big and small market data pooled together are $x_{0}$ = 0.2 (location) and   $\gamma$ = 2.0 (scale). As illustrated in {\bf Fig. \ref{Fig3.fig}}(A), the vast majority of observations are located around $x_{0}$. By way of example, in the full dataset (property listings sample size =133,668),   90\% (respectively, 99\%) of properties feature  $\vert \Delta P_{h} \vert \leq 10$\% (resp., $\vert \Delta P_{h} \vert \leq 40$\%). And in our final matched analysis  dataset (sample size = 57,414),   89\% of properties feature  $\vert \Delta P_{h} \vert \leq 10$\%. 
 
Nevertheless, the distribution  tails extend well beyond 10\%, indicating that fluctuations in this real estate asset class  are more  similar to the heavy-tailed  price fluctuation distributions observed for the equity asset class  \cite{mantegna1995scaling}. 
One explanation for the heavy tails is the large scale of real estate depreciation that can occur over the lifetime of ownership, balanced on the other side by relatively sudden appreciation attributable to renovations. Put another way, when a property enters the real-estate market, there is a rapid update in asset valuation that incorporates information that  had accrued over wide-ranging time scales.  This is of course not dissimilar from stock markets, where the periodic release of earnings and other news are rapidly absorbed into  stock prices  \cite{petersen2010market,petersen2010market}.

The price uncertainty  $U_{h}$ is a unique feature of the Zillow API data,  which also shows considerable variation, and is  narrowly centered around the 10\% level, but with significant right-skew. By way of comparison, consider the distribution $P(\Delta P_{h})$ calculated for properties listed for sale, for which we observe a systematic shift towards an excess frequency of $\Delta P_{h}>0$ values after 1/2020 relative to before. Conversely, in the case of $P(U_{h})$ we observe the opposite trend, signaling increased valuation confidence  after 1/2020 relative to before. Interestingly, in the case of rental properties, we observe no shift in $P(\Delta P_{h})$ comparing before and after, whereas the frequency of larger $U_{h}$ values post-1/2020 increases dramatically, possibly reflecting COVID-19 eviction moratorium policy rapidly implemented  in the US \cite{national2021rental,benfer2021eviction,gromis2022estimating}. See {\bf Fig. S4} for complementary distributions conditioned on market size, period and property type.\\

\vspace{0.1in}
\noindent{\bf Prominent shifts in real-estate valuation during COVID-19.} 
Using  the CA real estate market before 1/2020 as a comparative baseline,  {\bf Fig. \ref{Fig3.fig}}  shows that the post-1/2020 market   feature hallmarks of  a speculative bubble -- 
namely (a)  accelerated  valuation growth net of change in fundamentals and  (b) increased confidence in   excess valuation. Somewhat ironically, these characteristics may have emerged by way of  contagious  spreading of `irrational exuberance'  among market agents \cite{roehner2002patterns,glaeser2008housing,shiller2015irrational}.

One explanation for the  enhanced real-estate speculation derives from the global COVID-19 uncertainty shock, which  
muddled global expectations for   investment returns. 
This global shock resulted in a confounding and non-uniform impact on the public, as indicated by a diverging ``$K$-shaped'' recovery  in the US population \cite{dalton2021k}.
The shock was also followed by profound policy interventions, such as the sudden reduction of the US  federal funds target  rate taking the form of a long-lasting  financial-quake \cite{petersen2010quantitative,petersen2010market}, 
  which among other immediate effects promoted aggressive household borrowing that boosted home-purchasing power and home-improvement activity \cite{said2023person}. 
This also triggered a  sudden housing supply-demand imbalance exacerbated by the rapid expansion of remote work-from-home policy \cite{liu2021impact,mondragon2022housing}, in particular in the IT sector that is concentrated in the  Bay Area mega-region. While these factors primarily affect the house purchase market, they also augmented  uncertainty and speculation levels in the rental market, given the coincident increased demand for rent combined with sudden rent protection policy that together shifted  risk-levels for both tenants and rental property owners \cite{benfer2021eviction}. 

Combined, these factors are reflected by significant systematic shifts in the  characteristic levels of speculation ($ \Delta P_{h,m} $) and uncertainty ($U_{h,m}$)  across the entire range of $P_{h,m}$ -- for both small and big markets.  
Notably, we observe higher  average $\Delta P_{h,m}$    in small markets than in big markets, consistent with nationwide analysis  of the impact of state-level shutdowns on price changes in the months before and after their implementation, which were found to be mediated by differences in population and structural density between urban and rural markets \cite{d2022covid}. 

Compared with recent work analyzing the real estate market in southern CA that finds  a negative relation between price growth and price  \cite{landvoigt2015housing}, a relation that is consistent with  other asset classes such as firms and stocks \cite{buldyrev2020rise}, we instead observe an increasing trend in $\langle \Delta P_{h,m \rangle} \rangle$ with  $P_{h,m}$ after 1/2020, indicative of accelerated speculation -- see {\bf Fig. \ref{Fig3.fig}}(A). This shift is also readily apparent in the higher levels of price-growth variation ($\sigma[\Delta P_{h,m \rangle}]$) observed after 1/2020  -- see {\bf Fig. \ref{Fig3.fig}}(B). Again, this pattern deviates from the well-established decreasing size-variance relationship found for other asset classes \cite{mantegna1995scaling,stanley1996scaling,plerou1999similarities,mantegna1999introduction,riccaboni2008size,buldyrev2020rise}. Contrariwise, {\bf Fig. \ref{Fig3.fig}}(C,D) indicate a reduction in mean and standard deviation of price uncertainty after 1/2020, also consistent with the  conditions of a speculative bubble.
 \\

\noindent{\bf Quantifying the effect of COVID-19  on  speculative valuation in a CA real-estate market.} 
We use the rapid onset of the pandemic as an exogenous shock to uncertainty, which thereby facilitates estimating the  degree to which shifts in property valuation and valuation confidence during the pandemic were attributable to collective speculation.  
Our approach contributes to a growing body of quasi-experimental COVID-19 research in the social sciences  \cite{conley2021past}. 

As a consistency check,  we implemented two complementary quasi-experimental methods: (a) unit-level matching and (b) multivariate regression. Unit-level matching of individual properties  leverages the granularity of our data sample to estimate treatment effects manifesting at high spatiotemporal resolution. Instead, multivariate regression yields inferences based upon differences in group-level averages, with the notable advantage that additional regressors can be included in order to control for micro-level  (e.g., number of neighboring properties  listed for sale, $A_{h,m}$) and macro-level covariates (e.g., contemporaneous mortgage rates, $M_{m}$). 

Fundamental to both methods is identifying a counterfactual baseline to net out  differences  pre-existing before the pandemic.
To this end,  both approaches utilize  the rental market --  comprised of  properties that satisfy the same demand for housing, but were just  not available for sale and thus were neutral to contemporaneous speculation  -- as a counterfactual  baseline for comparison. Accordingly, both approaches rely on the parallel trend assumption between  on-market (denoted by ``For Sale'', $FS$) and off-market (``Rent'', $R$) property types, which we confirm and exhibit in {\bf Figure S7}.  

The logic underpinning this counterfactual approach is as follows. Whereas shifts in the valuation of on-market properties depend on shifts in the valuation of fundamentals in addition to market speculation,  shifts in the valuation of off-market properties  primarily reflect shifts in the valuation of fundamentals. Hence, we can estimate the impact of speculation on a given quantity $Y$ by way of a difference-in-difference (DiD) strategy denoted by 
\begin{eqnarray}  
\Delta \overline{\Delta}_{Y} := \overline \Delta_{\text{Y,FS}}-\overline \Delta_{\text{Y,R}} = \Delta (\text{Speculation})\ , 
\label{EQDiD}
\end{eqnarray}  
as illustrated in {\bf Fig. \ref{Fig2.fig}}(B).
For example, in the case of $Y={\Delta P}$ with  $\Delta F$ denoting the shift in price (${\Delta P}$) associated with fundamentals and $\Delta S$ denoting the shift in price associated with speculation, then the DiD corresponds to 
\begin{eqnarray}
   \Delta (\overline{\Delta}_{\Delta P}) &=& \overline{\Delta}_{\Delta P,\text{FS}} - \overline{\Delta}_{\Delta P,\text{R}} \nonumber \\ 
 = (\Delta P_{\text{Aft}}^{\text{FS}} &-& \Delta P_{\text{Bef}}^{\text{FS}}) - (\Delta P_{\text{Aft}}^{\text{R}} - \Delta P_{\text{Bef}}^{\text{R}}) \nonumber \nonumber \\ 
 = (\Delta F_{\text{Aft}}^{\text{FS}} &+& \Delta S_{\text{Aft}}^{\text{FS}} - \Delta F_{\text{Bef}}^{\text{FS}} - \Delta S_{\text{Bef}}^{\text{FS}}) - (\Delta F_{\text{Aft}}^{\text{R}}  - \Delta F_{\text{Bef}}^{\text{R}}) \nonumber \\ 
 =( \Delta S_{\text{Aft}}^{\text{FS}} &-& \Delta S_{\text{Bef}}^{\text{FS}} ) + ((\Delta F_{\text{Aft}}^{\text{FS}} - \Delta F_{\text{Aft}}^{\text{R}}) -  (\Delta F_{\text{Bef}}^{\text{FS}}  - \Delta F_{\text{Bef}}^{\text{R}}))  \nonumber \\ 
 = ( \Delta S_{\text{Aft}}^{\text{FS}} &-& \Delta S_{\text{Bef}}^{\text{FS}} ) = \Delta (\Delta S^{\text{FS}}) =  \Delta (\text{Speculation})
\end{eqnarray}
where the last line is follows if $(\Delta F_{\text{Aft}}^{\text{FS}} - \Delta F_{\text{Aft}}^{\text{R}}) = (\Delta F_{\text{Bef}}^{\text{FS}}  - \Delta F_{\text{Bef}}^{\text{R}})$, reflecting the assumption that there was no systematic shift in the value associated with changes in fundamentals-oriented valuation (e.g. renovation and maintenance costs) between the two property classes before and after 1/2020.
We apply this strategy to estimate the effect of the COVID-19 pandemic on two quantities that are sensitive to  uncertainty:  $Y=\Delta P_{h}$ and $Y=U_{h}$. 
Note  that  Eq. (\ref{EQDiD}), which we further specify in the following section, inherently  incorporates a temporal difference between the before and after 1/2020 periods. This second difference implies that the DiD  $\Delta \overline{\Delta}_{Y}$ is net of the baseline level of the market before 1/2020, meaning that this estimator quantifies the magnitude of price shifts specifically attributable to the speculation  in the CA real-estate market deriving from COVID-19 uncertainty.\\

{\it \noindent Method 1: Unit-level matching.} The quasi-experimental matching design implemented in this subsection does not conform to a traditional treatment-control setting    \cite{stuart2010matching}, as the pandemic was perniciously pervasive -- i.e.,  there is no untreated group in the period after 1/2020. 
Yet our approach still incorporates notable advantages of matching designs.
Foremost, this approach  accounts for unobserved covariates that are nonetheless correlated with the available matching variables. That is, while we do not explicitly incorporate house-specific features  -- such as  vicinity to shopping and schools, backyard size and other physical amenities such as a pool and garage,  these and many other variables are implicitly incorporated into each property valuation $P_{h,m}$ value, which we  use in the counterfactual matching stage. Moreover, by virtue of its design as a leading e-platform \cite{parker2016platform} that derives value by aggregating comprehensive and contemporaneous local and national house listings,  $P_{h,m}$ values are believed to be consistent and thus well-suited for the purpose of unit-level matching.

Our matching design also exploits the high geo-temporal resolution of the listing data to match properties listed after 1/2020 with  similar properties listed before 2020, thereby optimizing measurement precision  in the evaluation of market shifts due to  pandemic uncertainty. An advantage of this approach is  addressing the high degree of price and price change variation that exists even within a single region, as illustrated in {\bf Fig. \ref{Fig1.fig}}(C). To be specific, we account for unobserved unit-level features \cite{stuart2010matching} by strictly matching houses  according to three listing features: (a)  price strata, (b) calendar month, and (c) geographic location  -- variables that are only weakly related to the  variables of primary interest, namely   $\Delta P_{h,m}$ and $U_{h,m}$. 

We match on price strata by first calculating an intensive  variable  $Q_{c}(P_{h,m}) \in 1, 2 ...10$,  with 1 (respectively, 10) representing the lowest (highest) price decile  that is a specific to a particular city $c$ and before/after period. Assuming that potential buyers would be open to a range of house prices in excess of a single decile, we then  allow for matches  within $\pm 1$ decile group from $Q_{c}(P_{h,m})$. 
We constrained matches temporally by requiring matched houses from the same calendar month or 1 calendar month prior of the central house, which accounts for intra-year housing market  cycles. For example, if a property was listed in June, then we only accept  properties listed in May or June as candidate matches.
And we constrained matches geographically by requiring matched houses to be within a 0.5 mile (0.8 km) radius of the central house.

By way of example, {\bf Fig. \ref{Fig1.fig}}(G,H) illustrates the matching procedure using a property from  San Jose   listed after 1/2020, which also highlights the reduction in market supply after 1/2020 relative to before. Note that not all houses within the specified radius  are candidate matches because the  price variations in a single neighborhood can span several $Q_{c}(P_{h,m})$ strata. In {\bf Fig. \ref{Fig1.fig}}(G) we denote the set of matched houses in the same neighborhood of a given central house $h$ by $\{N_{h}\}_{\text{Bef}}$. 

 More specifically, for each property $h$  listed after 1/2020, we identify the match set $\{N_{h}\}_{\text{Bef}}$ from the pool of similar properties listed before 2020. 
We then construct a hypothetical property listed before 2020 that is very similar to $h$.  Ideally, the counterfactual property would be the same property $h$ using data sampled from before 2020. Unfortunately, the Zillow API only returns data contemporaneous to the data download date, and so we are unable to back-sample prior valuation data for any given property $h$. In order to overcome this challenge, a more sophisticated research design would need to identify a repeated sampling procedure to obtain a balanced Zillow estimates for the same set of properties over time, which was beyond the scope of our data collection capability, and is a limitation shared by most real-estate analyses using data (aggregated or not) for on-market properties. 

The characteristics of the counterfactual property are given by the average value $\langle Y \rangle_{\{N_{h}\}  \text{Bef}}$ calculated across the match set $\{N_{h}\}_{\text{Bef}}$, where $Y$ represents either $P_{h,m}$,  $\Delta P_{h,m}$ or $U_{h,m}$.
We then compute  for each $h$ the counterfactual difference 
\begin{eqnarray}  
\Delta_{Y,h} = Y_{h, \text{Aft}} - \langle Y \rangle_{\{N_{h}\}  \text{Bef}} \ ,
\label{MatchingEq}
\end{eqnarray}
 which estimates the shift in $Y$ associated with the two time periods. In a companion study, we perform a similar analysis by  instead matching first across property types within each time period, and then computing a temporal difference. This approach is more constrained by smaller $R$ sample sizes  for the period after 1/2020, yet we obtain  largely consistent results \cite{Zillow2_Petersen_2023}.
 
\begin{figure*}
\centering{\includegraphics[width=0.83\textwidth]{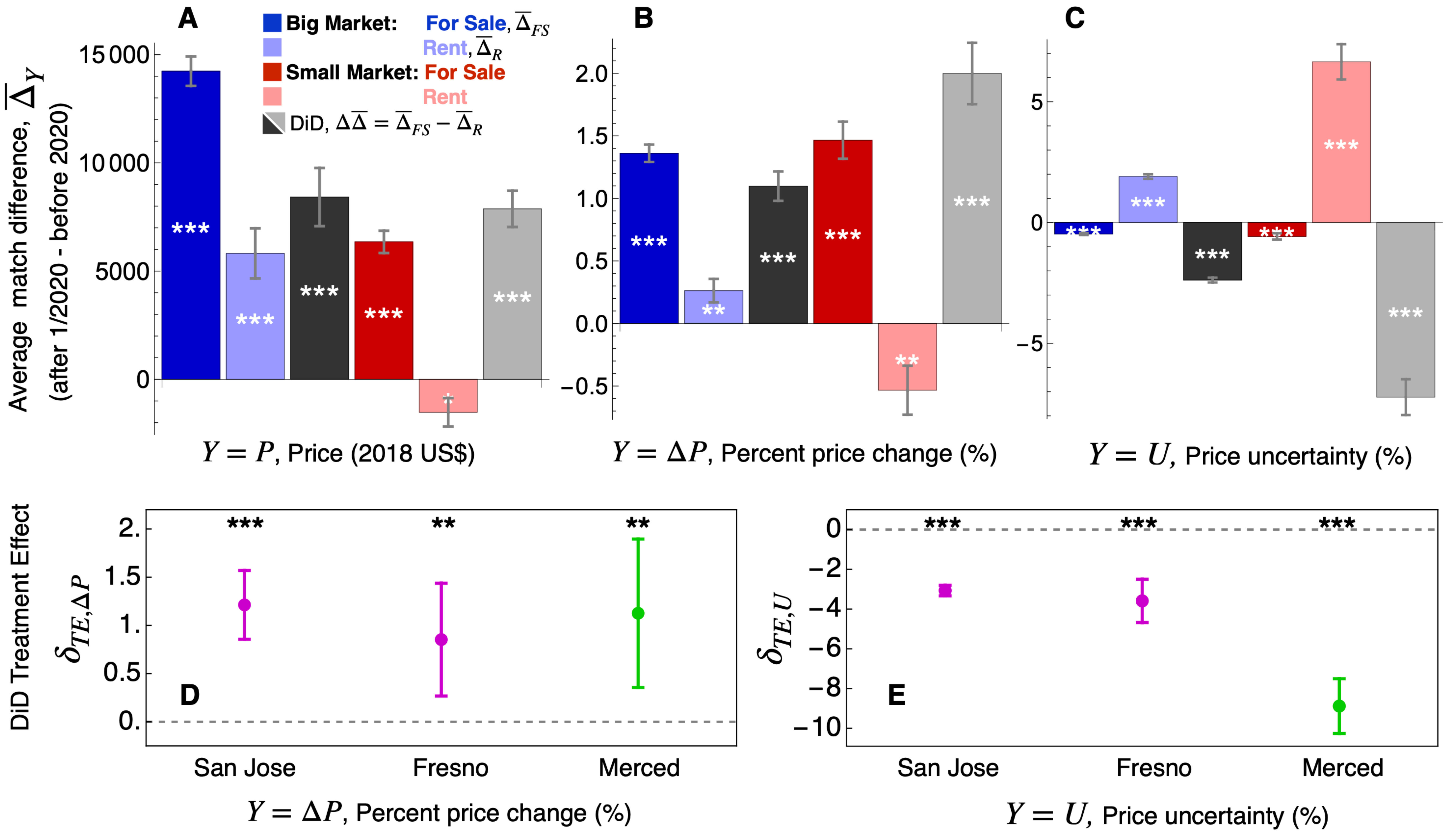}}
 \caption{  \label{Fig4.fig} {\bf Estimation of housing market valuation shifts attributable to COVID-19.}
(A-C) $\overline{\Delta}_{Y}$  is the distribution average of the unit-level difference $\Delta_{Y,h} = Y_{h, \text{Aft}} - \langle Y \rangle_{\{N_{h}\} \text{Bef}}$ calculated for the variable $Y$ across  properties   listed after 1/2020. The counterfactual baseline   $\langle Y\rangle_{\{N_{h}\}  \text{Bef}}$ is calculated using the set of  matched properties that were listed before 2020 (denoted by $\{N_{h}\}_{\text{Bef}}$).  
In this way, matching facilitates a more precise estimation of the impact of  COVID-19   on  individual properties. 
Error bars indicate the standard error of the mean and stars indicate the significance level of a T-Test for the likelihood of the  null hypothesis $\overline{\Delta}_{Y}=0$. 
Each gray  bar  represents the difference-in-difference $\Delta \overline{\Delta}_{Y} = \overline{\Delta}_{Y,FS} - \overline{\Delta}_{Y,R}$, which is an estimator for the effect of COVID-19 speculation on $Y$. Note that each market-level $\Delta \overline{\Delta}_{Y}$ is directly comparable and consistent with the corresponding city-level treatment effect $\delta_{TE,Y}$ shown in panel (D), where San Jose and Fresno are big markets, and Merced is a small market.
(A) The difference in the price  estimate  ($Y = P_{h,m}$; all values deflated to 1/2018 US\$)  shows the average price change for    listings after 1/2020.  
(B) The  difference in  price change  ($Y = \Delta P_{h,m}$) measures  shifts in  price valuations at high temporal resolution (30-day),  and shows that properties listed for sale had  excess price valuation relative to those listed for rent. 
(C) The  difference in   price uncertainty   ($Y =U_{h,m}$) is inversely related to  valuation  confidence. 
In the case of properties listed  for sale, we observe a 1-percentage point reduction in price-uncertainty, i.e.  higher valuation confidence; conversely, we observe drastic  price uncertainty increases for rental properties. 
(D,E)   Summary of the COVID-19 treatment effect $\delta_{TE,Y}$ on properties listed for sale, based upon results from a two-period difference-in-difference multivariate regression model. 
To summarize, average percent price change values increased between 0.85 and 1.21 percentage points, and price uncertainties declined between 3 and 9 percentage points, relative to the  baseline levels they plausibly would have maintained in the absence of  the  pandemic. Note that in both cases, this treatment effect corresponds to  properties listed for sale.
Error bars represent the 95\% confidence interval in each point estimate; full table of parameter estimates are reported in {\bf Tables S1-S2}. 
Significance levels indicated by the asterisks: * $p < 0.05$, ** $p < 0.01$, *** $p < 0.001$.}
\end{figure*}

 From the set of $\Delta_{Y,h}$ values collected for each region and property type, we then calculate the 
 average  difference 
 \begin{eqnarray}  
 \overline{\Delta}_{Y} = \langle \Delta_{Y,h} \rangle \ ,
 \end{eqnarray}
 where  we denote the  property type  in subscript, e.g. $\overline{\Delta}_{Y,FS}$ and $\overline{\Delta}_{Y,R}$.
 The impact of  the COVID-19 pandemic   on  the  variable $Y$ is then estimated according to the magnitude and statistical significance of $\overline{\Delta}_{Y}$. We evaluate the latter using a one-sample Student T-test to estimate the likelihood of the  null hypothesis $\overline{\Delta}_{Y}=0$ representing no pandemic effect. 
{\bf Figure \ref{Fig4.fig}}(A-C)  show the sign, magnitude and statistical difference of  $\overline{\Delta}_{Y}$ calculated for the three property-level variables $P_{h,m}$,  $\Delta P_{h,m}$ or $U_{h,m}$. See {\bf Fig. S5} for the distribution of individual $\Delta_{Y,h}$ values from which $\overline{\Delta}_{Y}$ are calculated; and  see  {\bf Fig. S6}  for   $\overline{\Delta}_{Y,c}$ calculated at the city level as a demonstration of robustness over down-scaled regions.   

Hence, the  difference in difference   $\Delta \overline{\Delta}_{Y}$  defined in  Eq. (\ref{EQDiD})  nets out the overall market shifts that may bias interpretation of $\overline{\Delta}_{Y,FS}$ when considered alone. What remains after subtracting our speculation-neutral baseline for comparison $\overline{\Delta}_{Y,R}$ is the excess impact attributable to speculation implicit in property sales. We evaluate the statistical significance of the null hypotheses $\Delta \overline{\Delta}_{Y}=0$  using the two-sample Student T-test with Welch correction that accounts for varying sample-size and variance between the $FS$ and $R$ samples. 

We begin by considering $Y$ values corresponding the absolute price change, which we report primarily for  the purpose of demonstrating that the magnitude of  price shifts we encountered are not incremental.   {\bf Figure \ref{Fig4.fig}}(A) shows   $\Delta \overline{\Delta}_{P} = \overline{\Delta}_{P,FS}-\overline{\Delta}_{P,R}$ of roughly 8,000 US\$ for both market sizes. This result indicates that the same property $h$ listed for sale is valued  8,000 US\$ more than if it was listed as available for rent, a result which is significant at the $p<0.001$  level. 

In what follows, we focus on the relative quantities $\Delta P_{h}$ and $U_{h}$, because intensive quantities (i.e., percentages) are more directly comparable, while also being less sensitive to the  matching variable $Q_{c}(P_{h,m})$. In particular,  {\bf Fig. \ref{Fig4.fig}}(B) indicates excess valuation growth over a 30-day period of   $\Delta \overline{\Delta}_{\Delta P} = \overline{\Delta}_{\Delta P,FS}-\overline{\Delta}_{\Delta P,R} =  1.36-0.26 =$ 1.1 percentage points for the average property in the big market, and $1.47-(-0.53)$ = 2.0 percentage points for the small market. Both DiD values are significant at the $p<0.001$  level. This result suggests that the valuation of the same property $h$  would  appreciate an additional 2\% percentage points more if it were listed for sale, as opposed to if it were instead listed as available for rent. In terms of the magnitude of this effect on properties listed for sale, the increase in $\Delta P_{h,m}$ is more than double the characteristic levels observed prior to the pandemic -- see  {\bf Fig. \ref{Fig3.fig}}(A).

Regarding percent price uncertainty, we  calculate  $\Delta \overline{\Delta}_{U} = \overline{\Delta}_{U,FS}-\overline{\Delta}_{U,R} =  -2.4$  percentage points for the big market,  and $\Delta \overline{\Delta}_{U}=$ -7.2 percentage points for the small market. Both DiD values are significant at the $p<0.001$ level.  This result suggests that the certainty in the valuation of a property  is higher if it were listed for sale than if it were listed as available for rent.

\begin{figure*}
\centering{\includegraphics[width=0.99\textwidth]{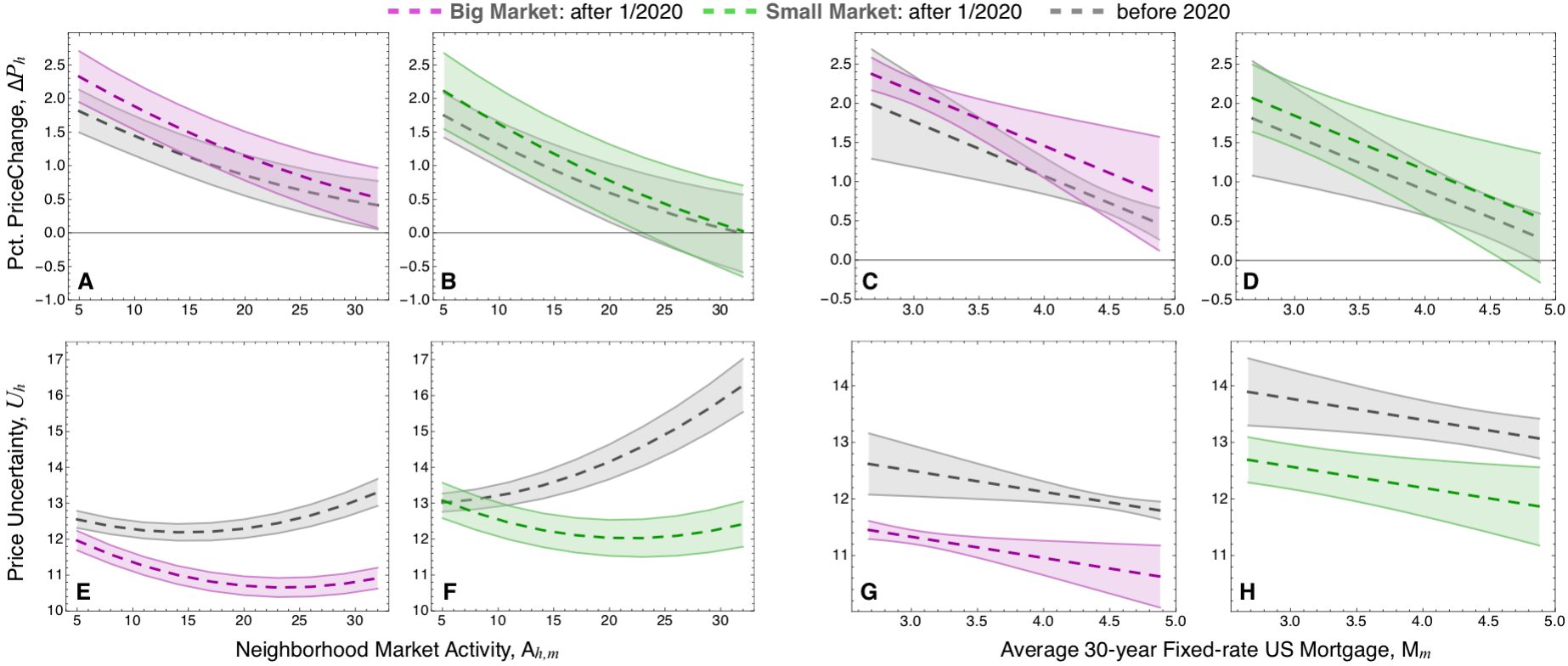}}
 \caption{  \label{Fig5.fig} {\bf Marginal effects of local market supply and  mortgage rate on price change and uncertainty.}  
(A,B) Predictions of the relationship  between the supply of alternative houses (defined as the number of  matched houses within the same period as the central house listing, $A_{h,m}$) and price change $\Delta {P}_{h}$. Positive shift in $\Delta {P}_{h}$ of roughly 0.5 percent after 1/2020 relative to before, which diminishes at higher levels of market supply for both small and big markets. 
(C,D) Predictions of the relationship between the average 30-year US Mortgage rate (Fixed rate, shown as percentage) and $\Delta {P}_{h}$. Positive shift on the order of 0.4 percent for both small and big markets. 
(E-H) Similar to panels (A-D) but showing the OLS model predictions for  price uncertainty. As expected, the uncertainty associated with COVID-19 is more clearly manifest in the market valuation uncertainty than the price dynamics. Counterintuitively,  the increased levels of uncertainty associated with the pandemic appear to have {\it reduced} uncertainty in price estimations, which points to the amplification of market speculation during this period of global stress. Shaded areas indicate 95\% confidence interval around the  predicted margins of response indicated by the dashed line. All marginal effects are scalculated using covariates maintained at their mean values.
 }
\end{figure*}

{\it \noindent Method 2: Multivariate regression.} We complement the  matching method with multiple regression, which affords estimating marginal relationships with temporal and spatial covariates. In what follows we implement a two-period difference-in-difference (DiD) model for  three regions (San Jose, Fresno, Merced) for which sufficient rental property data is available to serve as the before- and after-1/2020 control group. In short, we apply ordinary least squares (OLS) regression  using  STATA 13.0 software to estimate the following  model for   a specific region,
\begin{eqnarray}
Y_{h,m} = \delta_{TE} (I_{h,\text{ForSale}}\times T_{m}) + \vec{\beta} \cdot \vec{X} + \vec{\gamma} \cdot \vec{I} + \epsilon \ ,
\end{eqnarray}
 where $\vec{X}$ (respectively, $\vec{I}$)  represents a battery of continuous  (respectively, factor) controls, and the DiD interaction term $\delta_{TE} (I_{h,\text{ForSale}}\times T_{m})$ captures   the difference between the two property types (specified by the binary indicator variable $I_{h,\text{ForSale}}$) across the two  periods (specified by the binary indicator  $T_{m}$).  {\bf Figure S7} shows that the conditions of the DiD parallel trend assumption in the period before 2020 are sufficiently satisfied for both $\Delta P_{h,m}$ and $U_{h,m}$. And for additional  cross-validation,  see the study by  \cite{liu2021impact} analyzing repeated-transaction home price  data  within and across the 25 largest metropolitan statistical areas during the before-2020 period. And regarding the exclusion restriction on the treatment, one can verify this assumption by using Zillow.com to manually inspect properties listed for rent, and compare them to those that are listed for sale to see that there are no a priori systematic differences between the two property types. 

More specifically, we apply this canonical two-period  DiD specification to model two different dependent variables:  $Y \equiv \Delta P_{h,m}$ and  $Y \equiv U_{h,m}$. For each model we implement fixed-effects to  account for time-independent factors associated with the calendar month $m$ of the listing ($C_{m}$), and region-specific price strata  $Q_{c}(P_{h,m})$, where both quantities are encoded as categorial variables. Hence, the  treatment effect  $\delta_{TE}$ is the direct analog to $\Delta \overline{\Delta}_{Y}$, and estimates  the  excess shift in $Y$ attributable to  collective speculation deriving from   COVID-19 uncertainty.

In the first scenario where the dependent variable is the 30-day percent price change, the model specification is
\begin{eqnarray}
\Delta P_{h,m} &=&  \delta_{TE,\Delta P} (I_{h,\text{ForSale}}\times T_{m}) + \beta_{U}   (U_{h,m} \times I_{h,\text{ForSale}})  \\ \nonumber
 &+& \beta_{U^{2}}(U^{2}_{h,m} \times I_{h,\text{ForSale}})  +  \vec{\beta} \cdot \vec{X} + \vec{\gamma} \cdot \vec{I}  + \text{const.} +  \epsilon   \ , 
\label{M1a}	 
\end{eqnarray}
where the covariates are  $\vec{\beta} \cdot \vec{X} = \beta_{M} M_{m} + \beta_{A}(A_{h,m} \times I_{h,\text{ForSale}})  + \beta_{A^{2}}(A^{2}_{h,m} \times I_{h,\text{ForSale}})  +  \beta_{P}  \ln P_{h,m} + \beta_{P^{2}}  \ln^{2} P_{h,m}$ and the factor variables are  $\vec{\gamma} \cdot \vec{I}  = \gamma_{I} I_{h,\text{ForSale}} + \gamma_{T} T_{m} + \gamma_{Q}Q_{h} + \gamma_{C}C_{m} $.
The interaction between $I_{h,\text{ForSale}}$ and several  control variables  differentiate  responses conditional on property type.  
  Full model estimates are elaborated in {\bf Table S1}. 
  
Similarly,  in the second scenario where the dependent variable is the  percent price uncertainty, the model specification is
\begin{eqnarray}
 U_{h,m} &=&  \delta_{TE,U}(I_{h,\text{ForSale}}\times T_{m}) + \beta_{\Delta P}   (\Delta P_{h,m} \times I_{h,\text{ForSale}})  \\ \nonumber
 &+& \beta_{\Delta^{2} P} (\Delta^{2} P_{h,m} \times I_{h,\text{ForSale}})  +  \vec{\beta} \cdot \vec{X} + \vec{\gamma} \cdot \vec{I}  + \text{const.} +  \epsilon    \ .
\label{M2a}	 
\end{eqnarray} 
Full model estimates are elaborated in {\bf Table S2}

 {\bf Figure \ref{Fig4.fig}}(D) shows the  estimated treatment effect $\delta_{TE}$ for each model and city. Results indicate an excess 30-day percent price change of $\delta_{TE, \Delta P}=0.85$ (Fresno), 1.13 (Merced) and 1.21 (San Jose) percentage points. These values are consistent in sign, magnitude and statistical significance  with the corresponding market-level  DiD values $\Delta \overline{\Delta}_{\Delta P}$ estimated using the matching method.  Both methods indicate excess valuation, or higher valuations than there would have been in the absence of  COVID-19  market shock, which is consistent with prior theory of  housing-market speculation \cite{malpezzi2005role,glaeser2008housing}.

At the same time, we  observe declines in  price   uncertainties, corresponding to increases in valuation confidence, attributable to the pandemic:    $\delta_{TE,U}$ = -3.1 (San Jose), -3.6 (Fresno) and -8.9 (Merced) percentage points. 
As a robustness check, we confirm that each  point estimate $\delta_{TE,U}$ is consistent in sign, magnitude and statistical significance when compared with the corresponding market-level DiD values $\Delta \overline{\Delta}_{U}$ estimated using the matching method. See \cite{Zillow2_Petersen_2023} for additional consistency  check, where an analog  to Method 1 is applied, but instead  matches for-sale and rental properties within each  period and city, and then computes a pre-post difference as a DiD estimate.  

A clear limitation of our data sample is the lack of additional property-level feature data. As such,  
 unobserved factors may bias our $\delta_{TE,\Delta P}$ and $\delta_{TE,U}$ estimates, including  construction supply constraints \cite{yagi2021global,raetz2020hard}, the regulatory environment for affordable housing construction \cite{d2022covid}, shifts in demand for amenity density \cite{liu2021impact}, and remote-work and associated migration \cite{baum2019microgeography,mondragon2022housing}. \\

{\it \noindent Marginal effects of market supply and mortgage rates.} To further explore the relative impact   on price change and uncertainty, {\bf Fig. \ref{Fig5.fig}} shows the margins associated with (a) neighborhood market activity $A_{h,m}$, a micro-level indicator of housing supply measured as the number of potentially competing listings in the immediate vicinity of each listing; and (b) the average 30-year fixed-rate mortgage  $M_{m}$ obtained from Freddie Mac$^{\mbox{\scriptsize\textregistered}}$,  which is an inverse measure of homeowner borrowing power. 

The specification used to estimate these marginal effects is nearly identical to the DiD models described above. The main difference is we do not include the DiD term ($I_{h,\text{ForSale}}\times T_{m}$). Instead, this model includes an interaction $ S_{h} \times A_{h,m} \times T_{m} $ in order to quantify the marginal effect of neighborhood market activity $A_{h,m}$ associated with $Y=\{ \Delta P_{h,m}$ or $U_{h,m} \}$, while accounting for differences in period and market size. 
Full model estimates are elaborated in  {\bf Table S3}.

{\bf Figures \ref{Fig5.fig}}(A-D) provide an estimate of the semi-elasticity of price with supply, and are consistent in magnitude with  prior empirical work by \cite{saiz2010geographic} on the full elasticity of housing supply  conditioned by land development constraints. For example, an additional 10 local listings (i.e. $A_{h,m}$ shifting from 10 to 20) corresponds to a reduction in price change of roughly 0.6 (resp. 0.7) percentage points for the small (resp. big) market  before 2020; however, after 1/2020 this reduction increased in magnitude by roughly 0.1 percentage points for both markets as indicated by the increasingly steep slope after 1/2020.  

Another factor explaining  price gains during this period  are the lower interest rates that directly affect buyer purchasing power and builder construction costs \cite{HousingBoomMortgageRates}. The slope of the lines shown in {\bf Fig. \ref{Fig5.fig}}(C,D) provide an estimate of the  mortgage rate semi-elasticity, indicating a roughly 0.7 percent price increase for a 1 point reduction in  $M_{m}$, which is on the lower side but and consistent with estimation based upon a wide range of   approaches \cite{defusco2017interest}. The discrepancy may be attributable to the relatively low range of $M_{m}$ and relatively high monthly price changes encountered during our sample period. Note that the estimation for smaller (larger) interest rates for before (after) 1/2020 are extrapolations into out-of-sample $M_{m}$ regimes, as indicated by the larger standard errors indicated in the regression fit.

Another relevant analysis for comparison is one based upon the San Diego housing market from 1997-2008, which attributes higher price gains for houses at the lower end of the price distribution to cheaper credit \cite{landvoigt2015housing}. 
While we do not explicitly explore the interaction between $M_{m}$ and $\Delta P$ conditional on $P$,  we do not see evidence of the differential price gains by price segment over this period for big vis-a-vis small markets, as also indicated by {\bf Fig. \ref{Fig3.fig}}(A). 

{\bf Figure \ref{Fig5.fig}}(E-H) show analog response margins  associated with  price uncertainty $U_{h}$. 
For both big and small markets, uncertainty levels tempered after 1/2020 relative to before, corresponding to  higher levels of valuation confidence for the same levels of neighborhood supply. Counterintuitively, this result indicates more efficient  price discovery  \cite{barkham1996price}, despite greatly heightened socio-economic uncertainty.  Interestingly, the  informational signal captured by $A_{h,m}$ diminished during the pandemic in the small market, as indicated by the relatively flat profile in  {\bf Fig. \ref{Fig5.fig}}(F).

\vspace{-0.2in}
\section*{Discussion}
 \vspace{-0.2in}
The  rapid emergence of the global pandemic, followed by  pervasive mitigation policy, had broad yet uneven impacts across society \cite{bonaccorsi2020economic,bunn2021covid,dalton2021k,national2021rental,benfer2021eviction,gromis2022estimating,aleta2022quantifying}.
Against this backdrop,  we seek to contribute to the rich  literature emerging from this global crisis \cite{conley2021past} by utilizing this sudden uncertainty shock to analyze the collective dynamics of   real-estate price formation.

As all markets are highly correlated   systems  \cite{mantegna1999introduction,sornette2017stock,roehner2002patterns}, the COVID-19 pandemic perturbed the housing market in several critical ways.
First, the pandemic shifted  social interactions towards virtual modes, which increased the importance of online real-estate platforms as decision-making tools.
The  subsequent interruption to everyday life had immediate effects, as documented  in research showing that US counties featuring    stay-at-home orders also had higher property  sale prices \cite{d2022covid}. 
Other perturbations include   global  supply chain disruptions \cite{yagi2021global} that negatively impacted building costs and  exacerbated  supply inelasticity  \cite{baum2019microgeography}, two features that are central to the theory of emergent  housing bubbles \cite{malpezzi2005role,glaeser2008housing}. 
These supply factors were complemented by the expansion of remote-work options, which effectively increased the search radius of buyers, and decreased the overall demand for amenity density  \cite{liu2021impact}. 
Another pertinent factor in California are the pervasive regulations regarding real-estate development and new home construction \cite{raetz2020hard}. 

Viewed from  a longer perspective, the US real-estate market has been steadily transforming  since the   housing boom leading into the bust of 2007-2008.
In particular,  the growth  of the IT service economy  \cite{parker2016platform,maglio2019handbook} has  brought online real-estate platforms to ubiquity \cite{ZillowPopularityRank1}, with roughly 110 million distinct properties   tracked by Zillow Inc. \cite{ZillowCoverage}, corresponding to roughly 3 out of every 4  of the 142 million housing units tracked by the US Census Bureau in 2021. 
 In addition to updating  on-market and off-market property data, Zillow also calculates algorithmically consistent property valuations that are increasingly relevant to price formation in the US real-estate market.
 
The utility of such comprehensive and rapidly-updated market data extends far beyond active buyers and sellers. According to a recent  industry survey \cite{ZillowPorn}, 75\% of the respondents classify their time casually browsing real-estate platforms as an imagination outlet, with only 17\% claiming to search listings with  serious home-purchase motivations.
This statistic  suggests that, in addition to fundamental  shifts in supply and demand, the extreme levels of price growth  during the pandemic  may be attributable  to behavioral phenomena related to   heightened levels of life-course uncertainty, and an increased  prevalence of naive speculators that are important contributors to bubble formation \cite{SNLZillowPorn,hong2008advisors}.
Hence, inasmuch as  real-estate platform  service providers facilitate crowd-sourcing, browsing, and market-making, they also
 facilitate analyzing the dynamics of speculation at high resolution and vast scale. 

As such, this work consists of both methodological and  empirical contributions to the real-estate market literature. 
In order to address our three   research questions,  we first constructed  a high-resolution multi-region balanced panel comprised of individual property valuation estimates, which thereby facilitates inferential econometric analysis. Our main result is  estimating the excess price growth attributable to the COVID-19 pandemic  by way of two complementary econometric DiD approaches: unit-level matching and  multivariate regression.

Our property-level dataset combined with a pre-post model design leverages the systematic comparison of price estimates for properties listed for sale versus those listed for rent, the difference corresponding to the  effect of pandemic uncertainty on price speculation. Another  unique feature of our panel  is  its regional composition,  including both big (urban) and small (rural) real-estate markets.
In our first  DiD approach, we matched house listings based upon the set of available characteristics (listing month, price strata, longitude-latitude of the property)  to optimize around precision  in the calculation of the effect size  \cite{stuart2010matching}. 
In the second DiD approach, we implemented a canonical 2-period and 2-group model that incorporates additional covariates while also exploiting the different valuation and socio-economic features of renting versus buying that were exacerbated  during the pandemic.
Both approaches  yield consistent results, as summarized in {\bf Fig. \ref{Fig4.fig}}.
Because the 10 regions analyzed capture a relatively  wide variation in size, location and socio-economic backdrop, there is reason to  believe our results are generalizable to other US regions with housing markets similar to the Bay Area mega-region.\\

\noindent {\it Limitations:}  Our data and methods are characterized by various limitations. One limitation of our data sample is the lack of additional property-level feature data. As such,   unobserved factors may bias the $\delta_{TE,\Delta P}$ and $\delta_{TE,U}$ estimates produced by the multivariate regression method. Relevant omitted variables include  construction supply constraints \cite{yagi2021global,raetz2020hard}, the regulatory environment for affordable housing construction \cite{d2022covid}, shifts in demand for amenity density \cite{liu2021impact}, and remote-work and associated migration \cite{baum2019microgeography,mondragon2022housing}. These estimates may be further biased by spatial autocorrelation, which may call for more advanced  econometric methods employing spatial lag variables. However, we do note that our matching method accounts for time independent spatial autocorrelations, which are neutralized in the first difference applied in Eq. (\ref{MatchingEq}).

For this reason, we complemented the regression method by a   matching method, which constructs a hypothetical counterfactual property   according  to three matching factors: price, location and calendar listing month. In particular, we assume that the estimated price $P_{h,m}$ incorporates all the omitted variables in a consistent way. Hence, in matching properties according to price and location, we are able to   factor  out the missing idiosyncratic property details that contributed to each property's valuation.

Another notable limitation of our study is the inability to account for two complementary  demand-side factors, namely the shift towards remote work and the coincident emergence of online market intermediaries, or {\it iBuyers}. Regarding the former,  recent work shows that an increasing prevalence of remote work, and subsequent housing demand shifts associated with migration,   explains roughly half of the aggregate price changes over 2019-2021 \cite{mondragon2022housing}. 
Meanwhile, recent analysis on the emerging paradigm of instant-offer {\it iBuyer}  platforms finds that  the profitability of this emerging industry is highly impacted by valuation uncertainty  \cite{buchak2020intermediating}. 
Consequently, despite our analysis subsuming these factors, we are unable to cross-validate or contribute additional insights regarding their role in market speculation.

\vspace{-0.3in}
\section*{Conclusion}
 \vspace{-0.2in}
We analyzed the impact of the COVID-19 pandemic  shock using a property-level dataset including unique measures of uncertainty and  speculation. 
Despite the drastically increased  levels of uncertainty surrounding the scope and duration of the global pandemic, our results indicate a counterintuitive decrease in  property-level  price uncertainty ($U_{h,m}$).  At the same time, we employ two complementary methods to  estimate $\Delta \overline{\Delta}_{\Delta P}$ and $\delta_{TE,\Delta P}$, respectively, which quantify the excess price growth  attributable to heightened levels of pandemic speculation. Both methods yield consistent estimates,   on the order of 1\% per month excess price growth, i.e. above the levels of growth that would be expected in the absence of the pandemic, corresponding  to roughly +12.7 percentage points when integrated across an entire year. 
For context, this effect size   accounts for more than half of the actual  annual  growth  observed across these same  regions  in 2021. The coincidence of accelerating price growth and valuation confidence is a  hallmark of  a speculative bubble, which we found to  be stronger in the smaller housing markets, and likely reflects their greater susceptibility  to  sudden  supply contraction. 

Considered together, these results are harbingers of `irrational exuberance' \cite{shiller2015irrational} in  response to the sudden shock to long-term certainty  that augmented the dynamics and scale of collective speculation. 
These  findings, when contextualized against the backdrop of major life-course decision-making, are reconciled by behavioral theory regarding the persuasive power of uncertainty \cite{tormala2016role} and sudden unexpected interruptions \cite{kupor2015persuasion}. Considered in this light, while also accounting for  the magnitude of severity and surprise of this global shock, we speculate that the response to  COVID-19 uncertainty and subsequent daily life interruptions combined with the real-time inflow of market information collected by online real-estate platforms  may have contributed to collective herding behavior that is central to speculative  bubble formation in  complex socio-economic systems \cite{sornette2017stock,mantegna1999introduction,roehner2002patterns,hong2008advisors,glaeser2008housing,shiller2015irrational}.  \\

 \vspace{-0.1in}
\noindent {\bf Data availability}  All data analyzed here were sourced from the  open-access Zillow API \cite{ZillowAPI}. Anonymized data and code for reproducing the analysis will be made available at  Dryad upon publication \cite{DataDryad}. 



\newpage
\clearpage

\begin{widetext}

\beginsupplement 

\begin{center}
{\bf  \large Supplementary Information --  Appendix Text, Figures S1-S8 \& Tables S1-S3}
\end{center}

\bigskip
\bigskip
\bigskip

\noindent {\Large \bf Shift in house price estimates during COVID-19  \\ reveals  effect of crisis on collective speculation} 

\bigskip

\noindent {\bf \large  Alexander M. Petersen$^{1}$} 

\bigskip

\noindent $^{1}$Department of Management of Complex Systems, Ernest and Julio Gallo Management Program, School of Engineering, University of California, Merced, California 95343, USA \\

\begin{figure*}[hb!]
\centering{\includegraphics[width=0.69\textwidth]{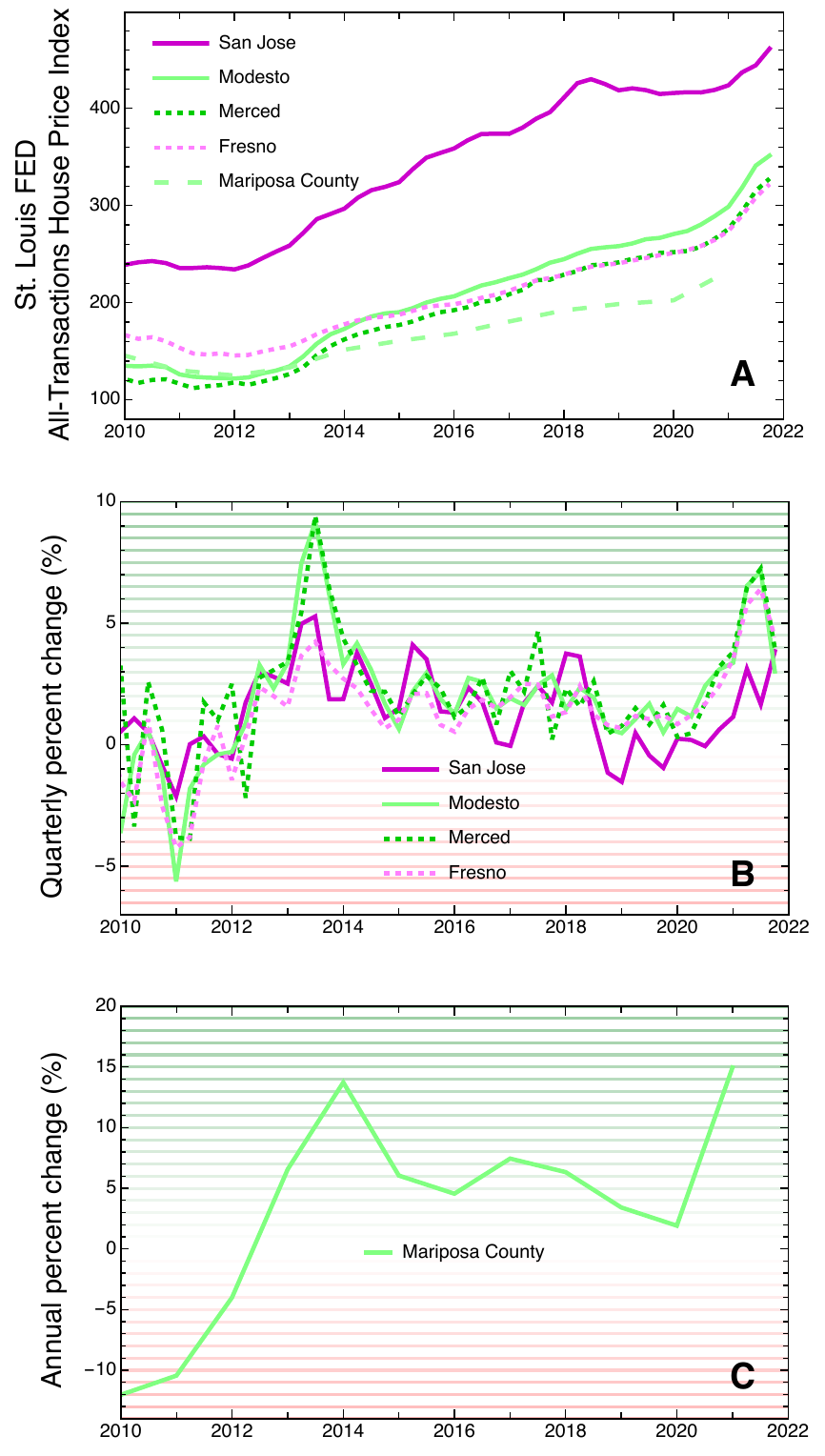}}
 \caption{   \label{FigS7.fig}  {\bf Quarterly price indices for several CA housing markets.} Aggregate region-level house price indices produced by the Federal Reserve Bank of St. Louis (quantities are ``Not Seasonally Adjusted'' and ``Estimated using sales prices and appraisal data.''). (A) Official Price Index data available for 
 \href{https://fred.stlouisfed.org/series/ATNHPIUS41940Q}{San Jose-Sunnyvale-Santa Clara};  
 \href{https://fred.stlouisfed.org/series/ATNHPIUS33700Q}{Modesto}; 
 \href{https://fred.stlouisfed.org/series/ATNHPIUS32900Q}{Merced}; 
 \href{https://fred.stlouisfed.org/series/ATNHPIUS23420Q}{Fresno}; 
 \href{https://fred.stlouisfed.org/series/ATNHPIUS06043A}{Mariposa County}.  (B,C) Two-period percent change, $\Delta_{t}(\%)  = 100(x_{t}-x_{t-1})/x_{t-1}$, calculated for the price indices ($x_{t}$) shown in panel A. Note that data for Mariposa are estimated at the annual frequency, and not the quarterly level as in the other cases, and because the range of values is substantially higher, we separated the panel for Mariposa County.
 }
\end{figure*}

\begin{figure*}
\centering{\includegraphics[width=0.85\textwidth]{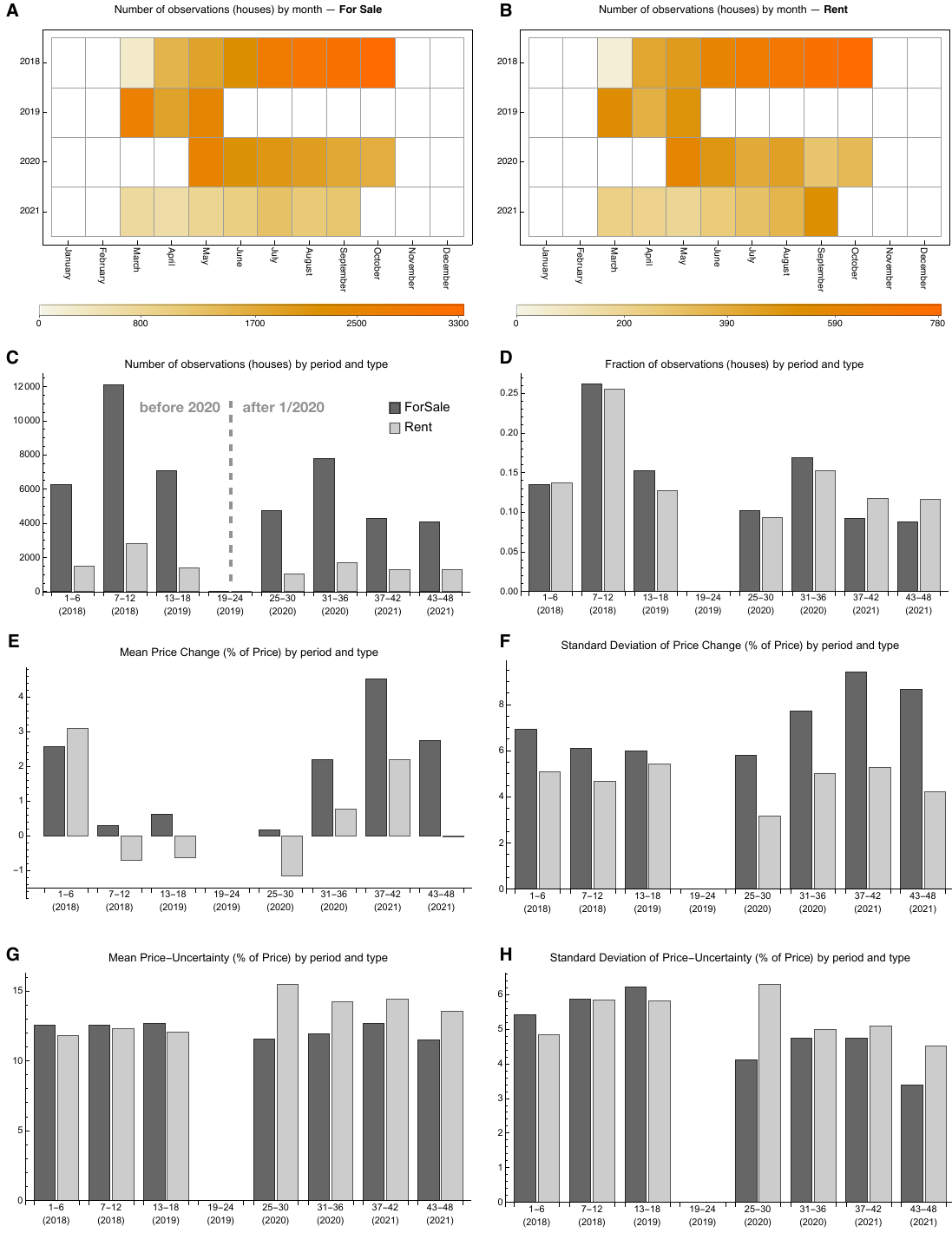}}
 \caption{  \label{FigS1.fig} {\bf Sample size and  summary  statistics grouped by period and property type.}. Observations separated into non-overlapping subsets according to  listing period and property types (For Sale and Rent). (A,B) Sample size by month. (C) Sample size as number of houses by period and type. (C) Sample size as fraction of  all houses belonging to a  period grouped by type, which shows common market size trends despite differences in absolute numbers. (E,F) Mean and standard deviation of ZEst house price, $P_{h}$. (G,H) Mean and standard deviation of ZEst house price uncertainty, $U_{h}$. There is a gap in data collection for month numbers 18 (June 2019) thru 27 (March 2020) due to changes in the Zillow website. Consequently, this restricted   our ability to obtain the house-level identifiers (ZPIDs) which were the principal input for the Zillow API for harvesting data; Data collection re-commenced in Feb. 2020, and since there is a 2-month padding to identify matched houses, this results in the data gap ending in May 2020. This data gap does not affect our ability to perform a pre/post analysis, as it falls principally during the housing market off-season of November thru February, which are summarily excluded from our analysis anyhow. Critical changes to the entire Zillow API platform in October 2021 haulted data collection entirely. }
\end{figure*}

\begin{figure*}
\centering{\includegraphics[width=0.99\textwidth]{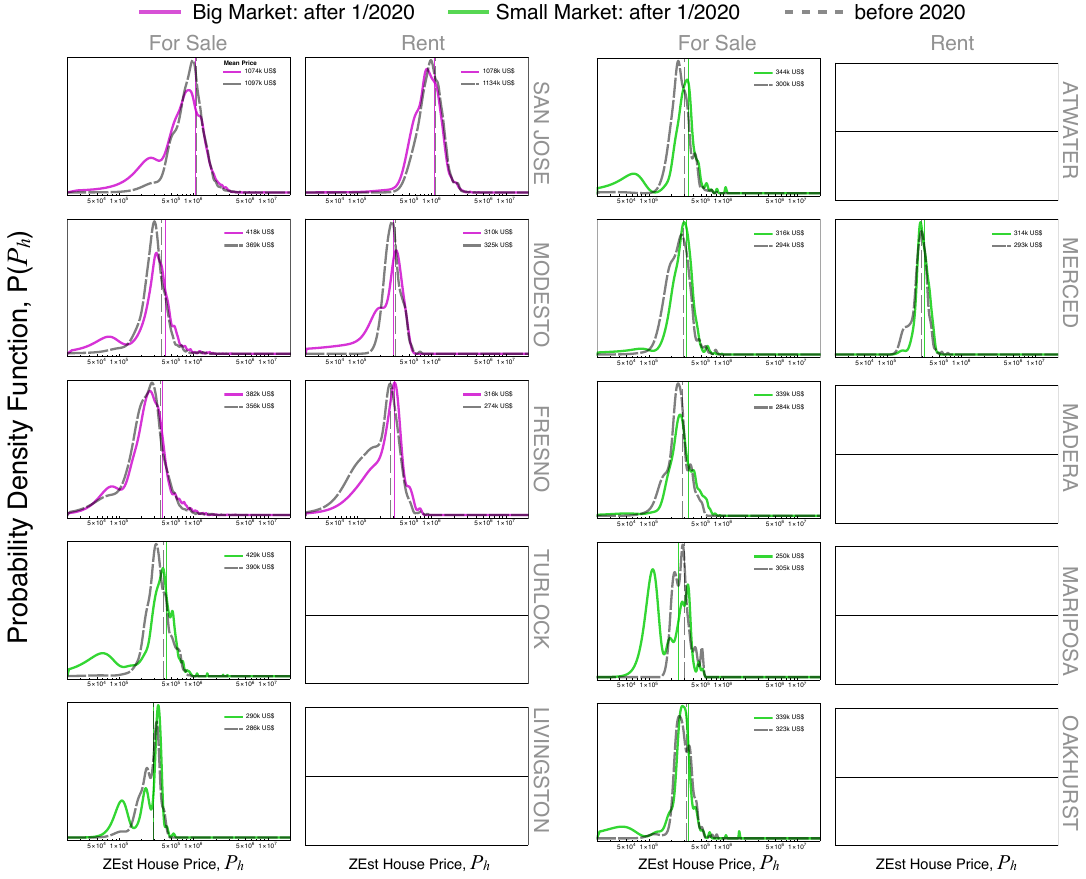}}
 \caption{  \label{FigS2.fig} {\bf Distribution of house price estimates grouped by city, period and property type.}
  Price distributions are organized by city in  two sets of columns according to housing market size (Big and Small). For each city we show the smooth kernel density estimate of the conditional price distribution, $P(P_{h} \vert \text{period, unit\ type})$, calculated according to  four non-overlapping data samples: for two periods  (before 2020 -- gray dashed curve; after 1/2020 -- colored solid curve) and two property types (For Sale and Rent).  To facilitate  comparing $P(P_{h} \vert \text{period, unit\ type})$ across period for a given property type,  vertical bars indicate the mean price value of the corresponding distribution. Note that all X-axes are shown on logarithmic scale, visually indicating that many $P(P_{h} \vert \text{period, unit\ type})$ are log-normal distributed. Also note that there was  only sufficient house rental data available through the Zillow API for 4 regions, with only one of these (Merced) belonging to the small market  group.  }
\end{figure*}

\begin{figure*}
\centering{\includegraphics[width=0.99\textwidth]{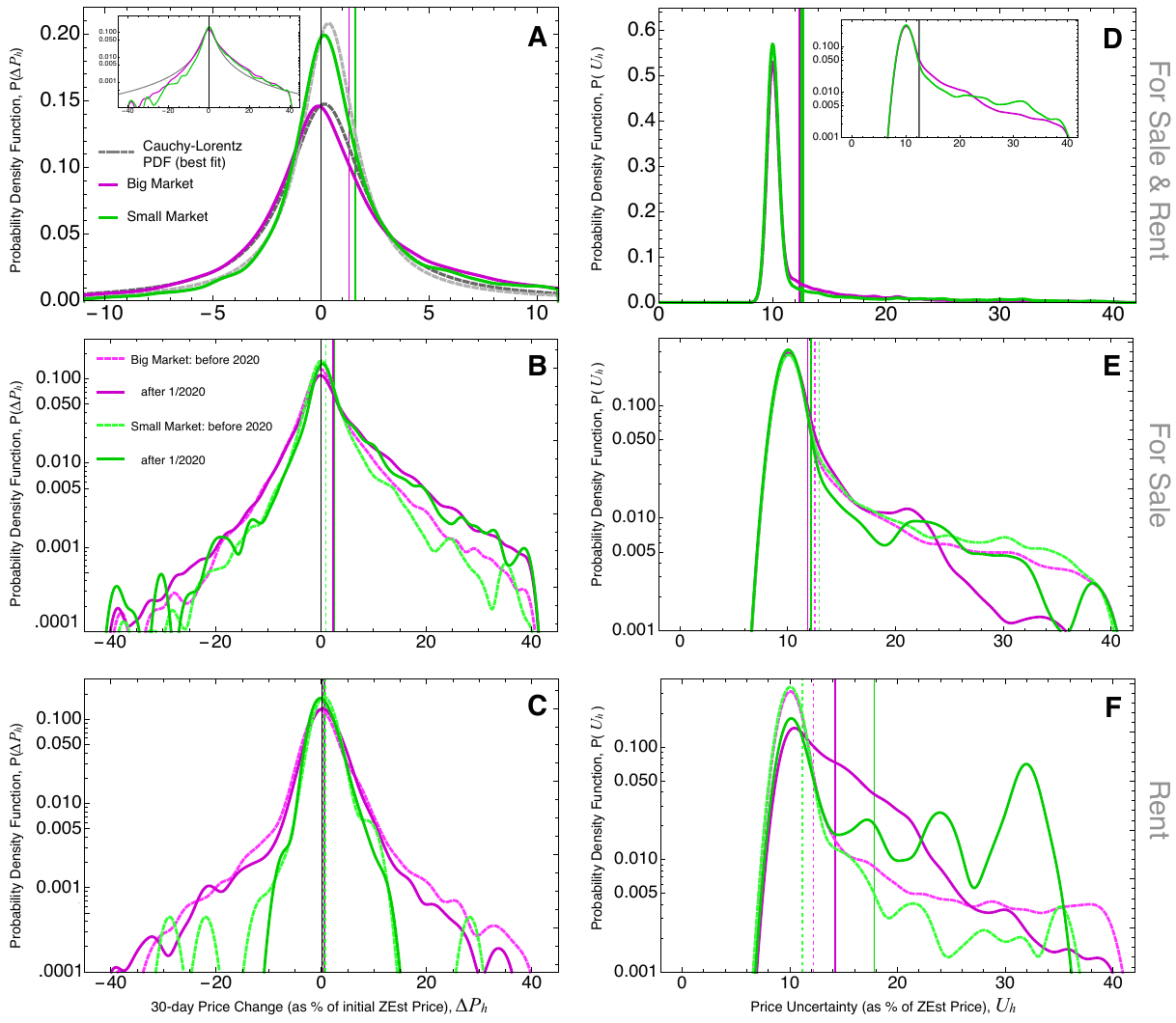}}
 \caption{ \label{FigS3.fig} {\bf Distribution of price change $\Delta P_{h,m}$ and uncertainty $U_{h,m}$ grouped by market size, period and property type.}
 Distributions of 30-day price change, $\Delta P_{h,m}$ (A-C) and price uncertainty, $U_{h,m}$  (D-F). For each data distribution we calculated the smooth kernel density estimate by collecting data into non-overlapping subsets based upon market size (Big and Small) and sampling period (before 2020 and after 1/2020).
(A) Aggregate price change distribution exhibits leptokurtic shape (i.e., broader than the benchmark Normal distribution), with the best-fit distribution model identified as the Cauchy-Lorentz probability density function (PDF) $P(x) \sim 1/x^{2}$ for $\vert (x - x_{0})/\gamma \vert \gg 1$. 
Each dashed line corresponds to the distributions by market size: For  properties in big markets, the location $x_{0}$ = 0.15 and  scale $\gamma$ = 2.2; similarly, for  properties in small markets,  $x_{0}$ = 0.34 and  scale $\gamma$ = 1.5. The bulk of the data are captured by $\vert \Delta P_{h} \vert \leq 10$, with  mean distribution values around 1.25 and 1.5\%  indicated by the solid vertical lines. 
The inset shows the distribution on log-linear axes, with the dashed line corresponding to the fit based upon both big and small market data pooled: $x_{0}$ = 0.2 and  scale $\gamma$ = 2.0. 
 (Inset) The price change distribution shown over the full range $\vert \Delta P_{h} \vert \leq 40$ \%. The empirical data distribution is  asymmetric, with empirical frequencies  in excess of  (less than) the best-fit Cauchy distribution for relatively large $\Delta P_{h} >0$ values ( $\Delta P_{h}<0$ values).
 (B) Price change distributions for houses listed for sale, by market and period, showing excess frequency for $\Delta P_{h}>0$ comparing after to before 2020, but not for $\Delta P_{h}<0$.  
 (C) Price change distributions for houses listed for rent, by market and period, where the main difference between the plots is associated with market size. 
 Comparing panels (B) and (C), the rent distribution is less leptokurtic in the bulk and also decays faster in both the positive and negative tails. 
 (D-F) Distributions of price uncertainty, $U_{h,m}$ indicate a skewed distribution closely centered around 10\% with mean values closer to 11\% in panels D and E which is dominated by properties listed for sale, and more variable in panel F which represents rental properties.  
}
\end{figure*}

\begin{figure*}
\centering{\includegraphics[width=0.99\textwidth]{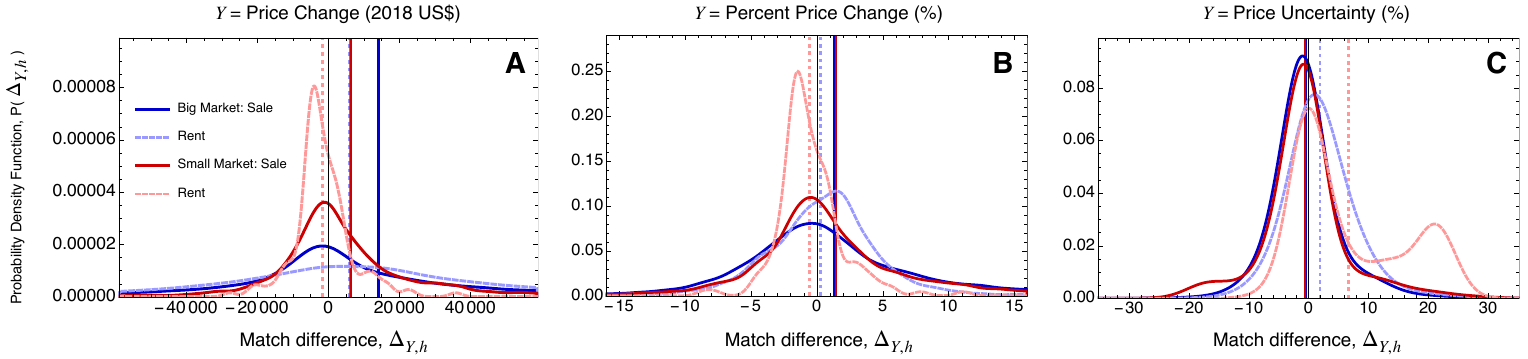}}
 \caption{   \label{FigS4.fig} {\bf Distributions of differences in matched-house price change and price uncertainty grouped by market size and property type.} 
Shown are the full distributions of match differences to supplement the mean match difference values ($\overline{\Delta}_{Y}$) reported in {\bf Fig. \ref{Fig4.fig}}. 
For each house listed after 1/2020 we calculate $\Delta_{h}$ between that house and the average value of $Y$ calculated across the set of  matched houses $\{N_{h}\}$  listed before 2020, i.e. $\Delta_{Y,h} = Y_{h} - \langle Y \rangle_{\{N_{h}\}}$, where the second term in the difference is the average value of $Y$ calculated across the set of matched houses that were listed before 2020.}
\end{figure*}

\begin{figure*}
\centering{\includegraphics[width=0.99\textwidth]{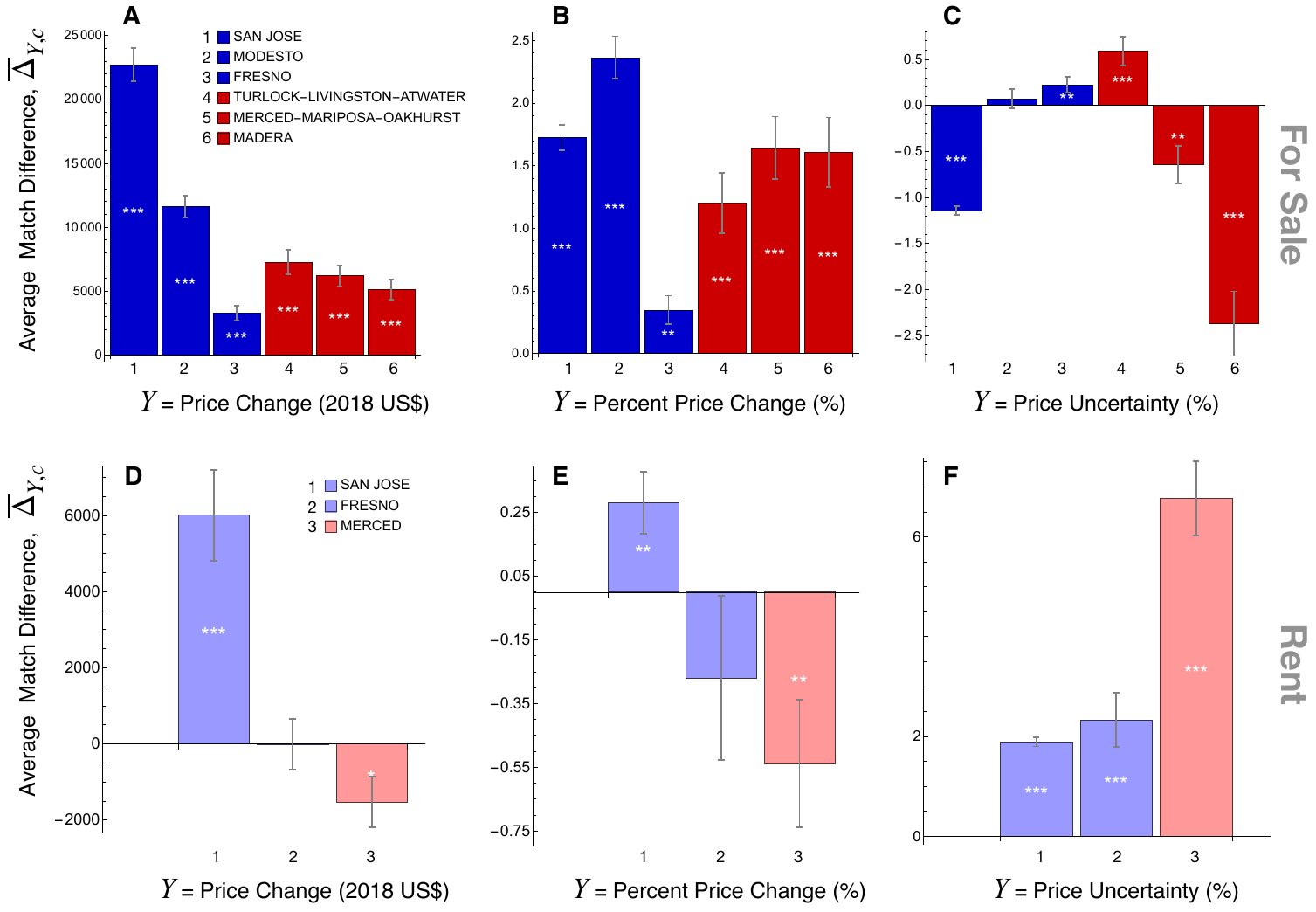}}
\caption{  \label{FigS5.fig} {\bf Estimation of housing market valuation shifts attributable to COVID-19 grouped by city.}
(A-C) Average match differences calculated for properties listed for sale. (D-F) Average match differences calculated for rental properties.}
\end{figure*}

\begin{figure*}
\centering{\includegraphics[width=0.99\textwidth]{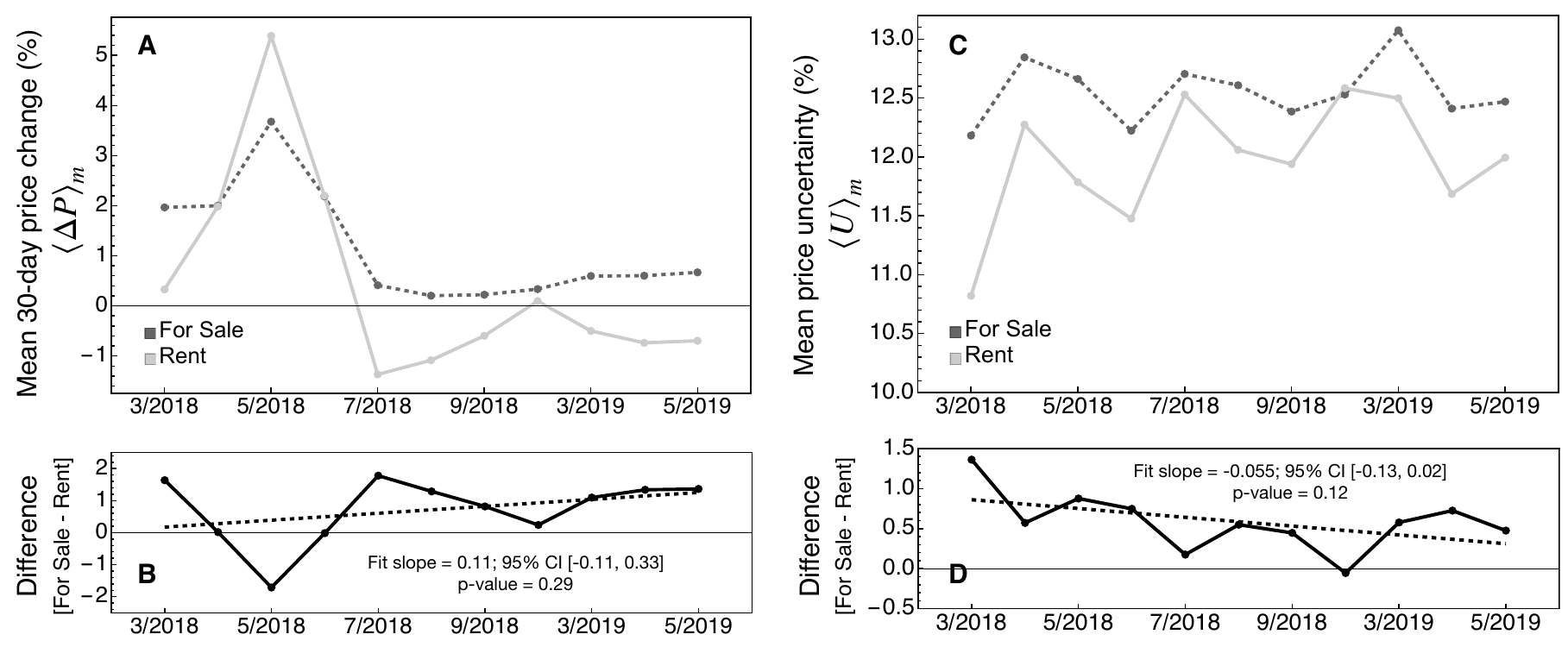}}
\caption{  \label{FigS8.fig} {\bf Test of the DiD parallel trend assumption.}
(A) Average percent price change by month $m$ for the 11 months in the data sample before 2020, $\langle \Delta P \rangle_{m}$, calculated for each property type. 
(B) Satisfactory parallel trends demonstrated by calculating the difference between the two curves in panel A and performing a linear trend OLS regression (parameter estimates shown in figure). Results indicate no significant trend. 
(C) Average price uncertainty by month $m$, $\langle U \rangle_{m}$, calculated for each property type.
(D) Satisfactory parallel trends demonstrated by calculating the difference between the two curves in panel C and performing a linear trend OLS regression (parameter estimates shown in figure). Results indicate no significant trend. }
\end{figure*}

\begin{figure*}
\centering{\includegraphics[width=0.99\textwidth]{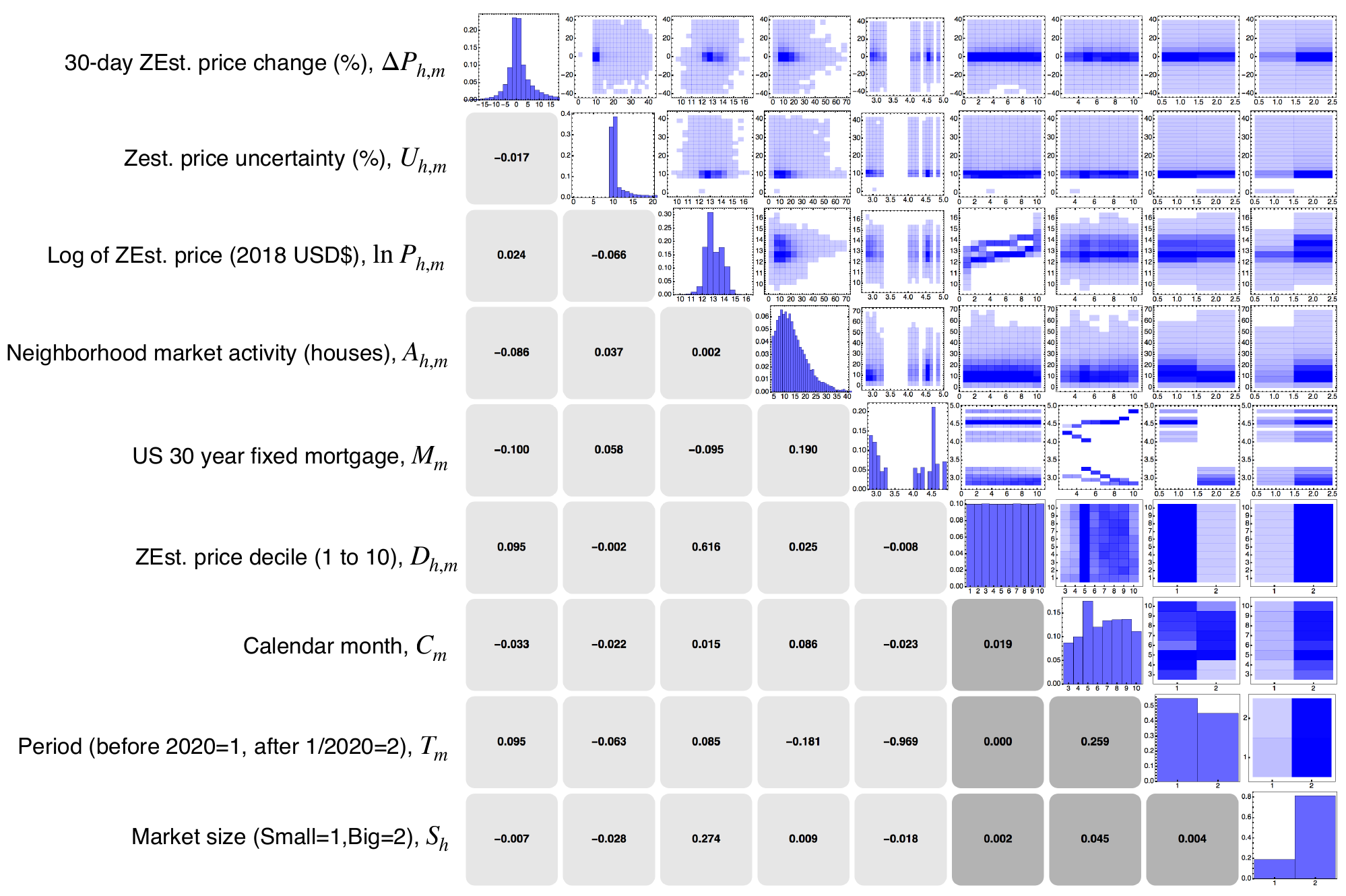}}
 \caption{   \label{FigS6.fig}  {\bf Cross-correlation and Descriptive statistics for regression model variables.} Upper-diagonal elements: bivariate histogram between row and column variables. Diagonal elements: histogram for variable indicated by the row/column labels. Lower-diagonal elements: bivariate cross-correlation coefficient: light-shaded squares indicate the Pearson's correlation coefficient between two variables that are both continuous measures; dark-shaded squares indicate the Cramer's V associate between two variables that are both  categorical.
 }
\end{figure*}

\clearpage
\newpage

\begin{table}[h!]
\centering
\caption{ {\bf Two-period Difference-in-Difference model with dependent variable   $\Delta P_{h,m}$}. Ordinary Least Squares (OLS) model implemented separately for the three regions with sufficient rental data, which serve as comparative DiD group, corresponding to  the $I_{h,\text{ForSale}}$ baseline: if $h$ is listed for sale, then $I_{h,\text{ForSale}} =0$ and $=1$ otherwise. 
Coefficients estimated with property type interaction are indicated by [For Sale]  and [Rent].
The coefficient $\delta_{TE,\Delta P}$ corresponds to the COVID-19 treatment effect on 30-day percent price changes, $\Delta P_{h,m}$, and visualized together in {\bf Fig. \ref{Fig4.fig}(D)}.  Note that $ T_{m}$ is the time period variable, taking the value 1 if the listing occurred after 1/2020, and 0 if before 2020.
Parameter estimate $p$-values are shown in parenthesis below each point estimate. OLS regression implemented in STATA 13 using ``reg''    calculated with robust standard errors.
Factor variables  included but not reported in the table below: Zest. price decile $D_{h,m}$, which ranges from 1 to 10; Dummy variable for calendar month, $C_{m}$, which ranges from 3 (March) to 10 (October) capturing intra-annual housing market cycle. See {\bf Fig. \ref{FigS6.fig}} for the cross-correlation matrix across the principal model covariates. }
\resizebox{0.85\columnwidth}{!}{ 
\def\sym#1{\ifmmode^{#1}\else\(^{#1}\)\fi}
\begin{tabular}{l*{3}{c}}
\hline\hline    
          &\multicolumn{1}{c}{}&\multicolumn{1}{c}{$Y= $ 30-day  Percent Price Change (\%), $\Delta P_{h,m}$}&\multicolumn{1}{c}{}\\
          &\multicolumn{1}{c}{\bf San Jose}&\multicolumn{1}{c}{\bf Fresno}&\multicolumn{1}{c}{\bf Merced}\\
\hline
Ave. US 30-yr. fixed mortgage rate, $\beta(M_{m})$ &     -0.0736         &      -0.306         &      -5.706\sym{***}\\
            &     (0.766)         &     (0.343)         &     (0.000)         \\
[1em]
Log of ZEst. price, $\beta(\ln P_{h,m}) $     &       22.17\sym{***}&       4.274         &       8.461         \\
            &     (0.000)         &     (0.312)         &     (0.517)         \\
[1em]
$\beta (\ln^{2} P_{h,m})$  &      -0.670\sym{**} &      -0.113         &      -0.309         \\
            &     (0.005)         &     (0.499)         &     (0.563)         \\
[1em]
Percent price uncertainty, $\beta(U_{h,m})$ [For Sale] &       0.326\sym{***}&      0.0398         &     -0.0591         \\
            &     (0.000)         &     (0.498)         &     (0.673)         \\
[1em]
Percent price uncertainty, $\beta(U_{h,m})$ [Rent]  &      -0.183\sym{*}  &      0.0520         &       0.465\sym{*}  \\
            &     (0.014)         &     (0.729)         &     (0.029)         \\
[1em]
$\beta(U_{h,m}^{2})$  [For Sale] &    -0.00850\sym{***}&    -0.00176         &     0.00298         \\
            &     (0.000)         &     (0.210)         &     (0.383)         \\
[1em]
$\beta(U_{h,m}^{2})$ [Rent] &     0.00572\sym{**} &   -0.000364         &     -0.0118\sym{*}  \\
            &     (0.003)         &     (0.921)         &     (0.034)         \\
[1em]
Market activity (neighboring houses), $\beta(A_{h,m})$  [For Sale] &     -0.0534\sym{*}  &      -0.129\sym{***}&     -0.0451         \\
            &     (0.011)         &     (0.000)         &     (0.303)         \\
[1em]
Market activity (neighboring houses), $\beta(A_{h,m})$ [Rent]  &     -0.0475         &       0.118         &     -0.0366         \\
            &     (0.218)         &     (0.471)         &     (0.384)         \\
[1em]
$\beta(A_{h,m}^{2})$ [For Sale] &    0.000389         &     0.00167\sym{**} &   -0.000515         \\
            &     (0.382)         &     (0.004)         &     (0.566)         \\
[1em]
$\beta(A_{h,m}^{2})$ [Rent] &     0.00367\sym{**} &    -0.00807         &    0.000797         \\
            &     (0.009)         &     (0.344)         &     (0.407)         \\
[1em]
After 1/2020 indicator, $\gamma(T_{m})$ &       0.882\sym{*}  &      -0.927         &      -8.668\sym{***}\\
            &     (0.032)         &     (0.085)         &     (0.000)         \\
[1em]
Property type indicator, $\gamma(I_{h,\text{ForSale}})$  &      -2.903\sym{**} &       2.973\sym{*}  &       4.794\sym{*}  \\
            &     (0.001)         &     (0.028)         &     (0.022)         \\
[1em]
{\bf Treatment effect}, $\delta_{TE,\Delta P}(I_{h,\text{ForSale}}\times T_{m})$  &       1.213\sym{***}&       0.853\sym{**} &       1.126\sym{**} \\
            &     (0.000)         &     (0.004)         &     (0.004)         \\
[1em]
Constant      &      -178.2\sym{***}&      -36.77         &      -32.13         \\
            &     (0.000)         &     (0.170)         &     (0.690)         \\
\hline
Fixed effect for price decile, $Q_{c(P_{h,m})}$ &    Y      &    Y    &    Y   \\
[1em]
Fixed effect for calendar month, $C_{m}$ &    Y      &    Y   &    Y    \\
\hline
\(N\)       &       25466         &       15495         &        3674         \\
adj. \(R^{2}\)&       0.060         &       0.018         &       0.059         \\
\hline\hline
\multicolumn{4}{l}{\footnotesize \textit{p}-values in parentheses}\\ 
\multicolumn{4}{l}{\footnotesize \sym{*} \(p<0.05\), \sym{**} \(p<0.01\), \sym{***} \(p<0.001\)}\\ 
\end{tabular}}
\label{TableS1.tab} 
\end{table}

\begin{table}[h!]
\centering
\caption{ {\bf Two-period Difference-in-Difference model  with dependent variable   $U_{h,m}$}. Ordinary Least Squares (OLS) model implemented separately for the three regions with sufficient rental data, which serve as comparative DiD group corresponding to the $I_{h,\text{ForSale}}$ baseline: if $h$ is listed for sale, then $I_{h,\text{ForSale}} =0$ and $=1$ otherwise. 
Coefficients estimated with property type interaction are indicated by [For Sale]  and [Rent].
The coefficient $\delta_{TE,U}$ corresponds to the COVID-19 treatment effect on the percent price uncertainty, $U_{h,m}$, and visualized together in {\bf Fig. \ref{Fig4.fig}(D)}.
Note that $ T_{m}$ is the time period variable, taking the value 1 if the listing occurred after 1/2020, and 0 if before 2020. 
Parameter estimate $p$-values are shown in parenthesis below each point estimate. OLS regression implemented in STATA 13 using ``reg'' calculated    with robust standard errors.
Factor variables  included but not reported in the table below: Zest. price decile $D_{h,m}$, which ranges from 1 to 10; Dummy variable for calendar month, $C_{m}$, which ranges from 3 (March) to 10 (October) capturing intra-annual housing market cycle. See {\bf Fig. \ref{FigS6.fig}} for the cross-correlation matrix across the principal model covariates. }
\resizebox{0.85\columnwidth}{!}{ 
\def\sym#1{\ifmmode^{#1}\else\(^{#1}\)\fi}
\begin{tabular}{l*{3}{c}}
\hline\hline    
          &\multicolumn{1}{c}{}&\multicolumn{1}{c}{$Y=$ \% Price Uncertainty, $U_{h,m}$}&\multicolumn{1}{c}{}\\
          &\multicolumn{1}{c}{\bf San Jose}&\multicolumn{1}{c}{\bf Fresno}&\multicolumn{1}{c}{\bf Merced}\\
\hline
Ave. US 30-yr. fixed mortgage rate, $\beta(M_{m})$ &       1.736\sym{***}&      -1.360\sym{***}&      -3.201\sym{***}\\
            &     (0.000)         &     (0.000)         &     (0.000)         \\
[1em]
Log of ZEst. price, $\beta(\ln P_{h,m}) $      &      -26.95\sym{***}&       3.414         &      -33.97\sym{**} \\
            &     (0.000)         &     (0.326)         &     (0.004)         \\
[1em]
$\beta (\ln^{2} P_{h,m})$  &       0.961\sym{***}&      -0.109         &       1.500\sym{**} \\
            &     (0.000)         &     (0.419)         &     (0.003)         \\
[1em]
Percent price change, $\Delta P_{h,m}$ [For Sale] &     -0.0166\sym{**} &     -0.0373\sym{***}&      0.0306         \\
            &     (0.009)         &     (0.001)         &     (0.090)         \\
[1em]
Percent price change, $\Delta P_{h,m}$ [Rent]  &      0.0390\sym{**} &       0.130         &     -0.0736         \\
            &     (0.005)         &     (0.148)         &     (0.411)         \\
[1em]
$\beta(\Delta^{2} P_{h,m})$  [For Sale]  &     0.00212\sym{***}&     0.00108\sym{*}  &     0.00165         \\
            &     (0.000)         &     (0.024)         &     (0.060)         \\
[1em]
$\beta(\Delta^{2} P_{h,m})$ [Rent] &     0.00895\sym{***}&      0.0186\sym{***}&      0.0171\sym{**} \\
            &     (0.000)         &     (0.001)         &     (0.009)         \\
[1em]
Market activity (neighboring houses), $\beta(A_{h,m})$  [For Sale] &     -0.0284\sym{*}  &      -0.155\sym{***}&      -0.244\sym{***}\\
            &     (0.013)         &     (0.000)         &     (0.000)         \\
[1em]
Market activity (neighboring houses), $\beta(A_{h,m})$ [Rent]  &      0.0926\sym{*}  &      -0.328         &       0.123         \\
            &     (0.023)         &     (0.213)         &     (0.244)         \\
[1em]
$\beta(A_{h,m}^{2})$ [For Sale]  &    0.000271         &     0.00621\sym{***}&     0.00769\sym{***}\\
            &     (0.263)         &     (0.000)         &     (0.000)         \\
[1em]
$\beta(A_{h,m}^{2})$ [Rent] &    -0.00171         &      0.0133         &    -0.00219         \\
            &     (0.270)         &     (0.367)         &     (0.427)         \\
[1em]
After 1/2020 indicator, $\gamma(T_{m})$ &       4.592\sym{***}&       1.469\sym{*}  &       2.236         \\
            &     (0.000)         &     (0.034)         &     (0.139)         \\
[1em]
Property type indicator, $\gamma(I_{h,\text{ForSale}})$  &       1.276\sym{***}&      -0.399         &       4.822\sym{***}\\
            &     (0.000)         &     (0.708)         &     (0.000)         \\
[1em]
{\bf Treatment effect}, $\delta_{TE,U}(I_{h,\text{ForSale}}\times T_{m})$  &      -3.067\sym{***}&      -3.588\sym{***}&      -8.882\sym{***}\\
            &     (0.000)         &     (0.000)         &     (0.000)         \\
[1em]
Constant     &       191.8\sym{***}&      -5.040         &       215.0\sym{**} \\
            &     (0.000)         &     (0.823)         &     (0.002)         \\
\hline
Fixed effect for price decile, $Q_{c(P_{h,m})}$ &    Y      &    Y    &    Y   \\
[1em]
Fixed effect for calendar month, $C_{m}$ &    Y      &    Y   &    Y    \\
\hline
\(N\)       &       25466         &       15495         &        3674         \\
adj. \(R^{2}\)&       0.075         &       0.081         &       0.139         \\
\hline\hline
\multicolumn{4}{l}{\footnotesize \textit{p}-values in parentheses}\\ 
\multicolumn{4}{l}{\footnotesize \sym{*} \(p<0.05\), \sym{**} \(p<0.01\), \sym{***} \(p<0.001\)}\\ 
\end{tabular}}
\label{TableS2.tab} 
\end{table}

\begin{table}[h!]
\centering
\caption{ {\bf  Aggregate model of properties listed `For Sale' with city fixed effects}. Parameter estimates for the model yielding  marginal effects plotted in {\bf Fig. \ref{Fig5.fig}}. $S_{h}$ is binary indicator variable coding the market size of each city (big or small).
Ordinary Least Squares (OLS) model implemented in STATA 13 using ``areg'' with city-level fixed effects, and calculated    with robust standard errors.  
Parameter estimate $p$-values are shown in parenthesis below each point estimate. }
\resizebox{0.62\columnwidth}{!}{ 
\def\sym#1{\ifmmode^{#1}\else\(^{#1}\)\fi}
\begin{tabular}{l*{2}{c}}
\hline\hline    
          &\multicolumn{1}{c}{$\Delta P_{h,m}$ (\%)}&\multicolumn{1}{c}{$U_{h,m}$ (\%)}\\
\hline
Ave. US 30-yr. fixed mortgage rate, $\beta(M_{m})$  &      -0.693\sym{***}&      -0.374\sym{*}  \\
            &     (0.000)         &     (0.012)         \\
[1em]
Log of ZEst. price, $\beta(\ln P_{h,m})$  &      -10.49\sym{***}&      -1.343         \\
            &     (0.000)         &     (0.254)         \\
[1em]
$\beta (\ln^{2} P_{h,m})$ &       0.462\sym{***}&      0.0422         \\
            &     (0.000)         &     (0.360)         \\
[1em]
Percent price uncertainty, $\beta(U_{h,m})$ &       0.114\sym{**} &                     \\
            &     (0.002)         &                     \\
[1em]
$\beta(U_{h,m}^{2})$  &    -0.00316\sym{***}&                     \\
            &     (0.001)         &                     \\
[1em]
Percent price change, $\beta(\Delta P_{h,m})$ &                     &     -0.0231\sym{***}\\
            &                     &     (0.000)         \\
[1em]
$\beta(\Delta^{2} P_{h,m})$ &                     &     0.00152\sym{***}\\
            &                     &     (0.000)         \\
[1em]
Market activity (neighboring houses),  $\beta(A_{h,m})$      &      -0.102\sym{***}&     -0.0180         \\
            &     (0.000)         &     (0.273)         \\
[1em]
$\beta(A_{h,m}^{2})$  &    0.000990\sym{***}&     0.00376\sym{***}\\
            &     (0.000)         &     (0.000)         \\
[1em]
After 1/2020 indicator, $\gamma(T_{m})$ &       0.425         &       0.787\sym{*}  \\
            &     (0.295)         &     (0.021)         \\
[1em]
$\beta(T_{m} \times A_{h,m})$ &     -0.0122         &      -0.145\sym{***}\\
            &     (0.430)         &     (0.000)         \\
[1em]
$\beta(S_{h} \times A_{h,m})$ &      0.0131         &     -0.0931\sym{***}\\
            &     (0.213)         &     (0.000)         \\
[1em]
$\beta(S_{h} \times T_{m})$  &       0.164         &      -1.043\sym{***}\\
            &     (0.598)         &     (0.000)         \\
[1em]
$\beta(S_{h} \times A_{h,m}  \times T_{m})$ &    -0.00286         &      0.0787\sym{***}\\
            &     (0.875)         &     (0.000)         \\
[1em]
Constant     &       61.14\sym{***}&       26.37\sym{***}\\
            &     (0.000)         &     (0.001)         \\
\hline
Fixed effect for price decile, $Q_{c(P_{h,m})}$ &    Y      &    Y      \\
[1em]
Fixed effect for calendar month, $C_{m}$ &    Y      &    Y      \\
\hline
\(N\)       &       46392         &       46392         \\
adj. \(R^{2}\)&       0.038         &       0.054         \\
\hline\hline
\multicolumn{3}{l}{\footnotesize \textit{p}-values in parentheses}\\ 
\multicolumn{3}{l}{\footnotesize \sym{*} \(p<0.05\), \sym{**} \(p<0.01\), \sym{***} \(p<0.001\)}\\ 
\end{tabular}}
\label{TableS3.tab} 
\end{table}

\end{widetext}

\end{document}